\documentclass[12pt, draftclsnofoot, onecolumn]{IEEEtran}
\usepackage{amsmath,amsthm,amsfonts,amscd,amssymb,graphics,dsfont}
\usepackage[mathscr]{eucal}
\usepackage{graphics,graphicx,multicol}
\usepackage{epsfig}
\usepackage{epstopdf}
\usepackage{subfigure}
\usepackage{stfloats}
\usepackage{latexsym}
\usepackage{amsfonts}
\usepackage{cite}
\usepackage{algorithm,algorithmic}
\usepackage{color,xcolor}
\usepackage{multirow}

\newtheorem{prop}{Proposition}

\newcommand{\qqed}{\hfill $\blacksquare$}

\begin{document}
\title{Learning Autonomy in Management of Wireless Random Networks}
\author{Hoon Lee,~\IEEEmembership{Member,~IEEE},
 Sang Hyun Lee,~\IEEEmembership{Member,~IEEE}, and Tony Q. S. Quek,~\IEEEmembership{Fellow,~IEEE}
\thanks{H. Lee is with the Department of Smart Robot Convergence and Application
Engineering and the Department of Information and Communications Engineering, Pukyong National University, Busan 48513, South Korea (e-mail: hlee@pknu.ac.kr).

S. H. Lee is with the School of Electrical Engineering, Korea University, Seoul 02841, South Korea (e-mail: sanghyunlee@korea.ac.kr).

T. Q. S. Quek is with the Information Systems Technology and Design Pillar, Singapore University of Technology and Design, Singapore 487372 (e-mail: tonyquek@sutd.edu.sg). }
}\maketitle 

\begin{abstract}
This paper presents a machine learning strategy that tackles a distributed optimization task in a wireless network with an arbitrary number of randomly interconnected nodes. Individual nodes decide their optimal states with distributed coordination among other nodes through randomly varying backhaul links. This poses a technical challenge in distributed universal optimization policy robust to a random topology of the wireless network, which has not been properly addressed by conventional deep neural networks (DNNs) with rigid structural configurations. We develop a flexible DNN formalism termed distributed message-passing neural network (DMPNN) with forward and backward computations independent of the network topology. A key enabler of this approach is an iterative message-sharing strategy through arbitrarily connected backhaul links. The DMPNN provides a convergent solution for iterative coordination by learning numerous random backhaul interactions. The DMPNN is investigated for various configurations of the power control in wireless networks, and intensive numerical results prove its universality and viability over conventional optimization and DNN approaches.
\end{abstract}


\section{Introduction}\label{sec:sec1}
The network management in next-generation wireless communication systems has encountered significant optimization challenges, including highly non-convex objectives, distributed coordination, large-scale system scalability, and robustness to topology changes. To circumvent these difficulties, intensive research activities have been conducted from perspectives of optimization \cite{Boyd:04,FRKschi:01,Boyd:10,MHong:16,YSun:17} and deep learning (DL) \cite{OShea:17,HLee:19a,HLee:20,WLee:18b,HSJang:19,WLee:18a,PKerret:18,Kim2018,HLee:19b,HSun:18}. Distributed optimization frameworks, such as dual decomposition \cite{Boyd:04}, belief propagation \cite{FRKschi:01}, and alternating direction method of multipliers \cite{Boyd:10} have been widely employed in distributed management for wireless networks. Recent efforts in the DL-based strategy open new opportunities for handling optimization formulations via deep neural networks (DNNs) \cite{WLee:18b,HSJang:19,WLee:18a}. The DL-based cooperation mechanisms are presented for distributed management, where individual nodes make local decisions via mutual coordinations exploiting network links \cite{PKerret:18,Kim2018,HLee:19b}.

However, the structural rigidity of existing approaches places a stumbling block to the extension to flexible networks, e.g., with time-varying properties in network dimension, node population, and node interaction. 
This becomes prominent even in the latest DL approaches \cite{OShea:17,HLee:19a,HLee:20,WLee:18b,HSJang:19,HSun:18,WLee:18a,PKerret:18,Kim2018,HLee:19b} since the corresponding DNNs should be trained independently in specific network configurations for universal use. This paper investigates a novel distributed DL strategy that applies universally to arbitrary network topologies. We put forth a universal DNN design so that forward and backward training computations accommodate the flexibility in the configuration.


DL techniques have recently addressed optimization tasks in end-to-end wireless transceiver design \cite{OShea:17,HLee:19a,HLee:20}, network resource allocation \cite{WLee:18b,HSJang:19,HSun:18,WLee:18a,FLiang:20,PKerret:18,Kim2018,HLee:19b}, and 5G/6G wireless communications \cite{CWang:20,CLuong:19,DLiu:20,AZhang:19}. The power control over interference channel (IFC) models has been intensively addressed in various ways \cite{HSun:18,WLee:18a,FLiang:20,PKerret:18,Kim2018,HLee:19b} since a seminal work in \cite{HSun:18} of  supervised learning to mimic a locally optimal weighted minimum mean-square-error (WMMSE) approach \cite{QShi:11}. Supervised learning techniques save significant computational complexity at the cost of the performance, while  unsupervised learning algorithms to optimize network utilities have also been developed in network applications \cite{WLee:18a,PKerret:18,Kim2018,HLee:19b}. An ensemble training technique is developed for improving unsupervised learning solutions via opportunistic choices for the best candidates \cite{FLiang:20}. Furthermore, DL-based approaches have addressed communication applications such as cognitive radios \cite{WLee:18b,HLee:19b} and non-orthogonal random access \cite{HSJang:19}.

Most DL-based works postulate a principle of centralism where a cloud server is responsible for collecting inputs from local nodes and evaluating a valid DNN output. By contrast, a {\em learning to cooperate} formalism \cite{AZhang:19,Gunduz2019} has been recently introduced to tackle the distributed network management in \cite{PKerret:18,Kim2018,HLee:19b}. The underlying policy is to decouple a node operation into two component DNN units: a message generator and a distributed optimizer. The message generator encodes locally available information into messages. The messages are subsequently transferred to nearby nodes over network links. The distributed optimizer combines incoming messages to determine the optimal state of the corresponding node. These units are trained offline in a centralized domain, while their inference is conducted {\em on the fly} in a decentralized manner.

Distributed wireless systems, e.g., internet-of-things (IoT) and wireless sensor networks, entail network configurations, such as network topology and node population, that are given arbitrarily and change gradually. However, existing distributed DL methods \cite{PKerret:18,Kim2018,HLee:19b} fail to grasp these random features since either the DNNs become ineligible to accommodate all possible candidates of networking setup for the limit of the capacity or their computations readily become prohibitively demanding. This gives rise to the necessity of a universal DL framework applicable to arbitrary network configurations.

We consider a distributed optimization problem over a random network. The distributed coordination allows nodes to share messages through backhaul links. Individual nodes produce the optimal solution based on messages along with local information. In reality, direct interactions among all nodes are not possible due to the absence of the backhaul links. Thus, the network model is inherently an undirected graph with randomly connected edges. With a graphical network model, an efficient DL computation structure of distributed optimizations is investigated. A single node aiming at distributed message-passing (DMP) inference is constructed with message generation, message reception, state update, and distributed decision. Each node generates a message dedicated to an adjacent node that is connected by a backhaul link. Subsequently, the received message is processed for a state variable update. The state variable is carefully designed to contain the information sufficient for the distributed identification of the optimal solution. The distributed decision is made by each node from the convergent state and messages at hand. Such a DMP iteration is repeated by sharing messages over a random backhaul graph. As opposed to one-shot message-passing approaches \cite{PKerret:18,Kim2018,HLee:19b}, where the states of individual nodes are determined using local observations and direct messages from adjacent nodes, the DMP inference enables multi-hop message transfers among nodes and allows indirect information sharing between unconnected nodes. For a versatile inference, the message reception consolidates a set function which does not depend on the number and the index of the input.

We propose a distributed message-passing neural network (DMPNN) framework to deploy the DMP inference. In particular, feedforward neural networks (FNNs) take care of message generation and distributed decision, while a recurrent neural network (RNN) updates the state so that its recursive structure reflects an iterative nature of the DMP operation. However, typical DNNs are inapt for adapting the message reception since the number of input messages varies with nodes according to the node degree and backhaul links.
To realize a universal DMPNN so that its computation is independent of the network topology, a viable message reception strategy is designed so that the corresponding DNN allows to take a set input \cite{MZaheer:17}. The DMPNN may be trained at a cloud server by observing numerous backhaul graphs with various edge and node configurations. For the distributed network operation, the DMPNN is deployed at individual nodes to accept the local information. It is tested with transmit power optimization problems over wireless networks and proves efficient with distributed optimization results.

The rest of the paper is organized as follows: Sec. \ref{sec:sec2} describes a system model for a random network  configuration and formulates a distributed optimization problem. Sec. \ref{sec:sec2.1} investigates special properties of the corresponding formulation, and Sec. \ref{sec:sec3} discusses the DMP inference as a universal solver. Sec. \ref{sec:sec5} presents the DMPNN framework along with training and inference strategies. Numerical results are presented in Sec. \ref{sec:sec6}, and the paper is concluded in Sec. \ref{sec:sec7}.

\textit{Notations:} We represent matrices, vectors, and scalar quantities in uppercase boldface letters, lowercase boldface letters, and normal letters, respectively. Also, sets of $m$-by-$n$ real-valued matrices and $m$-dimensional real-valued vectors are denoted as $\mathbb{R}^{m\times n}$ and $\mathbb{R}^{m}$, respectively. All zero column vector of length $m$ is denoted by $\mathbf{0}_{m}$. 

\section{System Model}\label{sec:sec2}
\subsection{Interaction Models}
\begin{figure}
\centering
\includegraphics[width=.8\linewidth]{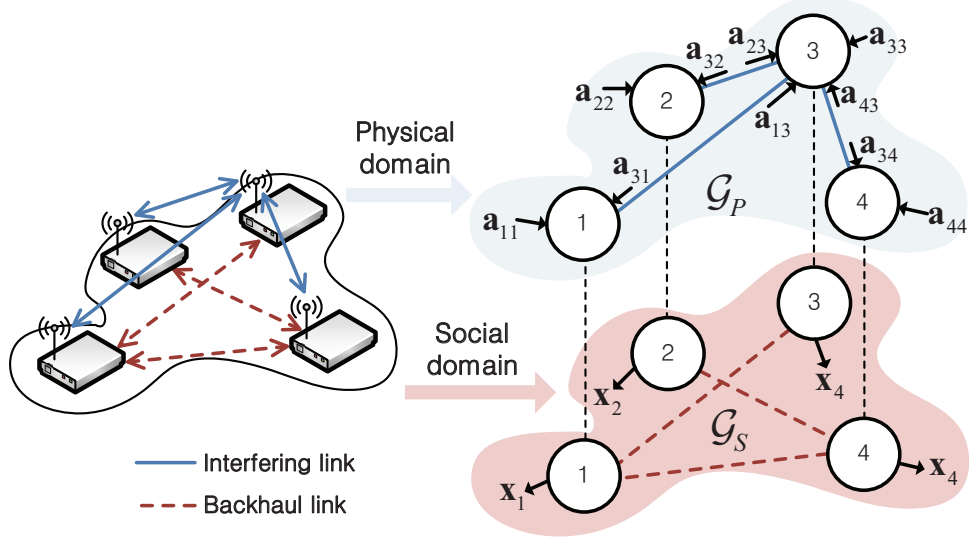}
\caption{A general multiplex network setup with $N=4$ nodes where physical and cooperative interactions are different.}
\label{fig:fig1}
\end{figure}

Consider a multiplex network \cite{GBianconi:18} with the set of $N$ nodes $\mathcal{V}\triangleq\{1,\ldots,N\}$ as illustrated in Fig.~\ref{fig:fig1}. In wireless networks, access points, base stations, and mobile devices become nodes. Their simultaneous message transmission may interfere with each other. Interactions among wireless nodes are modeled in two different domains: a physical domain and a social domain. In a multiplex network, two component graphs associated with the corresponding domains, respectively, are defined over the same set of nodes, and no direct connection exists between nodes in different domains. The physical domain captures a physical interaction environment, e.g., the interference relationship among different cells, while the social domain characterizes a logical communication link configuration, i.e., the backhaul infrastructure.

The interactions in both domains are represented with undirected graph models. A {\em physical graph} $\mathcal{G}_{P}=(\mathcal{V},\mathcal{E}_{P})$ describes interactions in the physical domain with respect to node set $\mathcal{V}$ and edge set $\mathcal{E}_{P}$. 
An edge $(i,j) \in \mathcal{E}_{P}$ 
is associated with two nodes $i$ and $j$ interfering with each other.
Let $\mathcal{N}_{P}(i)\triangleq\{j:(i,j)\in\mathcal{E}_{P}\}$ be a {\em physical neighborhood} that represents the set of all nodes adjacent to node $i$ in the physical domain.
Node $i\in\mathcal{V}$ produces a $X$-dimensional solution vector $\mathbf{x}_{i}$ based on an $A_{i}$-dimensional observation vector $\mathbf{a}_{i}$. It is the concatenation of the internal information $\mathbf{a}_{ii}$ of length $K_{1}$ and the collection of external data vector $\mathbf{a}_{ji}$ of length $K_{2}$ originating from each node $j\in \mathcal{N}_{P}(i)$ in the physical neighborhood, i.e., $\mathbf{a}_{i}\triangleq\{\mathbf{a}_{ii}\}\cup\{\mathbf{a}_{ji}:j\in \mathcal{N}_{P}(i)\}$.\footnote{A physical domain graph model can be generalized with a complete graph that contains all possible edges by setting $\mathbf{a}_{ji}$ and $\mathbf{a}_{ij}$ to null vectors for edge $(i,j)$ absent from $\mathcal{G}_{P}$.} The dimensions of $K_{1}$ and $K_{2}$ are predetermined based on system parameters such as the number of transmit antennas and the user population. Thus, the dimension of $\mathbf{a}_{i}$ is $A_{i}=K_{1}+K_{2}|\mathcal{N}_{P}(i)|$. Since $\mathbf{a}_{ji}$ is regarded as an edge attribute, it includes any information that node $i$ passively senses from node $j$. In particular, 
$\mathbf{a}_{ii}$ corresponds to the user channel state information that cell $i$ in a multi-cell network obtains, if designated as node $i$, whereas $\mathbf{a}_{ji}$ is interpreted as the interference originating from adjacent cell $j \in \mathcal{N}_{P}(i)$.

The backhaul links are not necessarily directly related to their physical interactions. The corresponding social interactions are normally independent of the physical domain, and a {\em social graph} $\mathcal{G}_{S}=(\mathcal{V},\mathcal{E}_{S})$ characterizes the node cooperation. Thus, edge set $\mathcal{E}_{S}$ represents the backhaul configuration, e.g., edge $(i,j) \in \mathcal{E}_{S}$ indicates that nodes $i$ and $j$ can exchange information via a backhaul link. In practical applications such as IoT and sensor networks, a communication link within a node pair often becomes unavailable although those two nodes have a mutually interfering effect in the physical domain. 
Let $\mathcal{N}_{S}(i)\triangleq\{j:(i,j)\in\mathcal{E}_{S}\}$ be a {\em social neighborhood} corresponding to the set of all nodes adjacent to node $i$ in $\mathcal{G}_{S}$. Node $i$ can forward the information regarding observation vector $\mathbf{a}_{i}$ and solution $\mathbf{x}_{i}$ to social neighbors in $\mathcal{N}_{S}(i)$ through connected links. Nodes can have active interactions of message exchanges only with the social neighborhood. In this configuration, a distributed strategy desires each node $i$ to determine own solution $\mathbf{x}_{i}$ using $\mathbf{a}_{i}$ independently to optimize the network-wide utility. 

\subsection{Universal Formulation}
To assess the efficiency of the optimization solution, 
a (possibly nonconvex) network utility function $f(\mathbf{a},\mathbf{x})$ is defined with respect to the global observation vector $\mathbf{a}\triangleq\{\mathbf{a}_{i}: i\in\mathcal{V}\}$ and the collection of local decisions $\mathbf{x}\triangleq\{\mathbf{x}_{i}: i\in\mathcal{V}\}$. 
To cope with random communication topology $\mathcal{G}_{S}$ and arbitrary node population $N$, a universal design of a distributed optimization strategy is necessary. Node $i$ aims at identifying its solution $\mathbf{x}_{i}$ such that individual node contributions cooperatively maximize the network utility averaged over the observation $\mathbf{a}$ and the social graph $\mathcal{G}_{S}$. The random effects of backhaul cooperation $\mathcal{E}_{S}$ and node deployment $\mathcal{V}$ with arbitrary $N$ are incorporated to the objective function. The corresponding optimization is formulated as
\begin{align*}
(P):\quad &\max_{\mathbf{x}}~\mathbb{E}_{\mathbf{a},\mathcal{G}_{S}}[f(\mathbf{a},\mathbf{x})]\quad \text{subject to}\ \mathbf{x}_{i}\in\mathcal{X},
\end{align*}
where $\mathcal{X}$ indicates a feasible solution space.

We desire to solve (P) so that node $i$ determines its solution $\mathbf{x}_{i}$ through distributed cooperations with the social neighborhood $\mathcal{N}_{S}(i)$, but not with the physical neighborhood $\mathcal{N}_{P}(i)$. Viable solution candidates for (P) include a combined design of information sharing policy over arbitrary backhaul connections and distributed decision strategy based on the local observation. In contrast to traditional distributed optimization problems that have a typical assumption of $\mathcal{G}_{S}=\mathcal{G}_{P}$, solving (P) for a network configuration of arbitrary population $N$ with $\mathcal{G}_{S}\neq\mathcal{G}_{P}$ is a highly challenging task that normally becomes intractable with existing techniques. The discrepancy between physical and social graphs prevents the direct sharing of the physical observations $\{\mathbf{a}_{i}\}$ through social domain connections. Individual nodes need to induce suitable sufficient statistics for handling (P) and interaction strategies over dynamic network topology. Furthermore, wireless nodes are responsible for the self-organizing management computation to process the attributes dedicated to distributed nodes, such as local observations $\{\mathbf{a}_{i}\}$ and local connection topology $\mathcal{N}_{P}(i)$ and $\mathcal{N}_{S}(i)$. However, existing information delivery and routing protocols may not properly address such challenges since they typically lack local statistics sharing policies and resort to the knowledge of global topology $\mathcal{G}_{S}$.

We propose to use a DL framework for tackling such technical challenges. Several latest works have investigated DL-based strategies to address nonconvex optimization tasks in wireless network management \cite{WLee:18b,HSJang:19,HSun:18,WLee:18a,FLiang:20,PKerret:18,Kim2018,HLee:19b}.
These approaches essentially rely on fully-connected DNN layers that construct rigid structures of fixed input and output dimensions. For this reason, the scaling-up and adaption of the techniques to variable network population $N$ necessarily involve training multiple instances of DNN, often resulting in a prohibitive amount of computations. To resolve such challenges, one valid alternative is a convolutional neural network (CNN) that replaces computations by convolutional operations with parameters tuned for input instances. This has been recently applied for association tasks in wireless networks \cite{WCui:19}. The wireless node deployment is envisioned in a two-dimensional (2D) image with each pixel value representing the node population at a grid area. This 2D input allows convolutional layers to maintain the invariance to topology and population. At the final step of the scheduling solution, however, the assignment of node strategies still requires a fixed output layer as large as the node population. A fully convolutional network (FCN) constructed purely with convolutional layers \cite{fcn} may handle this difficulty. The computations in the FCN are made independent both of input and output dimensions. To this end, a lattice graph where a vertex (or a pixel) is connected to nearby vertices in the Euclidean space is considered. By contrast, a class of optimization tasks in (P) is defined over vector spaces associated with a pair of independent graphs $\mathcal{G}_{P}$ and $\mathcal{G}_{S}$. Physical and social interactions are not directly captured in geometric node configurations. Thus, CNN and FCN techniques are not eligible for networking applications under arbitrary connection patterns.

To overcome this issue, a graph neural network (GNN) has been applied in wireless communication systems \cite{YYang:20,MLee:21,YShen:19,MEisen:19,AChowdhury:21}. This is an extension of a CNN to graph domains where graph convolutional operations aggregate interconnected node inputs. Parameters shared by nodes among subgraph combinations lead to flexible structures realized by a stack of graph filter layers. This approach lends itself to scalable solutions of massive identification applications such as multi-antenna channel estimation \cite{YYang:20}, link scheduling \cite{MLee:21}, and resource management \cite{YShen:19,MEisen:19,AChowdhury:21}. However, a decentralized realization based on this framework has not been properly addressed especially via backhaul coordination mechanisms, i.e., in \cite{YYang:20,MLee:21,YShen:19}, centralized data collection steps are necessary for the global network information.  Although a few distributed GNN implementations are presented \cite{MEisen:19,AChowdhury:21}, handcrafted interaction policies request fixed topology configurations. Thus, all node pairs guaranteed by dedicated connections \cite{AChowdhury:21} prevents the variation of social graph $\mathcal{G}_{S}$. Furthermore, an identical structure of the corresponding physical and social domains undermines an adaptation to a practical networking setup where interfering nodes are not aligned with communicating ones. The detailed descriptions with existing GNN approaches are made and compared in Sec. \ref{sec:sec5b}. Therefore, it is essential to develop a novel distributed DL strategy for (P) flexibly configurable for an arbitrary value of $N$ and random instances of $\mathcal{G}_{P}$ and $\mathcal{G}_{S}$.


\section{Network Management Strategy}\label{sec:sec2.1}
We first study some special properties of the optimal solution for (P) that prove useful in introducing a flexible DNN structure. We consider the utility function invariant with the permutation of the underlying graph. Let $\pi:\mathcal{V}\rightarrow\mathcal{V}$ be a permutation that changes a node index in a graph. Superscript $\pi$ represents the permuted version of the corresponding quantity with $\pi$, and $\pi(i)$ indicates the permuted index of node $i$. A permuted physical graph $\mathcal{G}_{P}^{\pi}\triangleq(\mathcal{V}^{\pi},\mathcal{E}_{P}^{\pi})$ is defined with $\mathcal{V}^{\pi}=\{\pi(1),\cdots,\pi(N)\}$ and $\mathcal{E}_{P}^{\pi}\subseteq\mathcal{V}^{\pi}\times\mathcal{V}^{\pi}$. Likewise, $\mathcal{G}_{S}^{\pi}\triangleq(\mathcal{V}^{\pi},\mathcal{E}_{S}^{\pi})$ denotes the corresponding permuted social graph. Let us define $\mathbf{a}^{\pi}\triangleq\{\mathbf{a}_{i}^{\pi}:i\in\mathcal{V}^{\pi}\}=\{\mathbf{a}_{\pi(i)}^{\pi}:i\in\mathcal{V}\}$ and $\mathbf{x}^{\pi}\triangleq\{\mathbf{x}_{i}^{\pi}: i\in\mathcal{V}^{\pi}\}=\{\mathbf{x}_{\pi(i)}^{\pi}: i\in\mathcal{V}\}$. A permutation-invariant utility $f(\mathbf{a},\mathbf{x})$ is a function satisfying
$f(\mathbf{a},\mathbf{x})=f(\mathbf{a}^{\pi},\mathbf{x}^{\pi})$ 
with respect to the graph permutation $\mathcal{G}_{P}^{\pi}$.\footnote{Although all notions in this work apply to any type of the utility function, this work focuses on the permutation-invariant utility since it leads to intriguing results.} For example, wireless resource management problems with uniform node and resource types are considered in a class of the permutation-invariant problems. The solution of such a class is of a particular structure described in the following statement.
\begin{prop}\label{prop:prop0}
Let $\mathbf{x}_{i}~(i\in\mathcal{V})$ and $\mathbf{x}^{\pi}_{j}~(j\in\mathcal{V}^{\pi})$ be the optimal distributed solutions of (P) for two graph pairs $(\mathcal{G}_{P},\mathcal{G}_{S})$ and  $(\mathcal{G}_{P}^{\pi},\mathcal{G}_{S}^{\pi})$ rearranged by permutation $\pi$, respectively.
It holds that
\begin{align}
\mathbf{x}_{i}=\mathbf{x}^{\pi}_{\pi(i)},\ \forall i\in\mathcal{V}.\label{eq:prop0}
\end{align}
\end{prop}
\begin{IEEEproof}
The proof is provided in Appendix \ref{app:appAA}.
\end{IEEEproof}

Proposition \ref{prop:prop0} states that the optimal state $\mathbf{x}_{i}$ of each node $i$ equals the optimal solution $\mathbf{x}_{\pi(i)}^{\pi}$ over new graph $(\mathcal{G}_{P}^{\pi},\mathcal{G}_{S}^{\pi})$ configured by permutation $\pi$. In other words, the optimal distributed solution remains permutation-equivariant such that the node ordering does not affect the solution computation results. This is reminiscent of the share of identical optimization strategies among all nodes, rather than node-specific calculation rules. 
It is a crucial factor for a universal optimization framework that adapts to an arbitrary social graph with randomly varying configurations. 
For universal operation with diverse configurations of hyperparameters relevant to various network topology,
the training computation proceeds with their subnetwork configurations with a small number of nodes, instead of the entire network of a whole  population, which normally requires prohibitively large computations and data sets. This is valid because, even in a large-scale network, only a set of local nodes forms the neighborhood of individual nodes, on average, and individual nodes learn to determine their states using interactions only among them.

Proposition \ref{prop:prop0} enables to determine a single computation strategy that can be shared by all nodes. The solution of (P) still relies on graph-specific calculations since it may be different for graphs that are not permutations of each other. As a result, an efficient DNN-based solver for (P) possesses an permutation-equivariant architecture in \eqref{eq:prop0} and simultaneously captures physical and social connections as input features. This motivates to develop the DMPNN framework to compute the DMP inference presented in the following sections.

\subsection{Network Management Applications}\label{sec:sec3a}



We address applications 
in the resource management of interference channels (IFCs) where $N$ transmitter-receiver pairs communicate with the same radio resources. Transmitters are designated as nodes in (P) responsible for solving distributed power control problems. A physical graph $\mathcal{G}_{P}$ represents interfering relationship among receivers, i.e., transmitter $i$ interferes with receiver $j\in\mathcal{N}_{P}(i)$. By contrast, backhaul connections among  transmitters are characterized by a social graph $\mathcal{G}_{S}$. Let $a_{ji}$ be a channel gain of the link between transmitter $j$ and receiver $i$. Thus, $a_{ii}$ and $a_{ji}$ are both scalars, i.e., $K_{1}=K_{2}=1$. The interfering link channel $a_{ji}$ is available at transmitter $i$ which is responsible for estimating the information signal transmitted from transmitter $j$ via uplink channel feedback \cite{HLee:19b}, whereas the local information $a_{ii}$ is a channel gain of the link between transmitter $i$ and its receiver. Denoting $x_{i}$ by transmit power level at transmitter $i$, achievable rate of transmit-receiver pair $i$ is expressed as
\begin{align}
r_{i}(\mathbf{a},\mathbf{x})=\log\bigg(1+\frac{a_{ii}x_{i}}{1+\sum_{j\in\mathcal{N}_{P}(i)}a_{ji}x_{j}}\bigg).
\end{align}
Two relevant objectives of sum rate and minimum rate can be considered for the network management. The sum rate maximization is formulated as
\begin{align*}
(P1):\quad&\max_{\mathbf{x}}~\mathbb{E}_{\mathbf{a},\mathcal{G}_{S}}\bigg[\sum_{i\in\mathcal{V}}
    r_{i}(\mathbf{a},\mathbf{x})\bigg]\\
     & \text{subject to} ~ x_{i}\in[0,P],\ i\in\mathcal{V},
\end{align*}
while the minimum rate counterpart is given by
\begin{align*}
(P2):\quad&\max_{\mathbf{x}}~\mathbb{E}_{\mathbf{a},\mathcal{G}_{S}}\bigg[\min_{i\in\mathcal{V}}
    r_{i}(\mathbf{a},\mathbf{x})\bigg]\\
    &\text{subject to} ~ x_{i}\in[0,P],\ i\in\mathcal{V}.
\end{align*}
where $P$ is the maximum power constraint.
By symmetry, both objectives are readily found to be permutation invariant.


\section{Distributed Message Passing Inference}\label{sec:sec3}
\begin{figure*}
\centering
    \subfigure[Computation of node $i$]{
        \includegraphics[width=.28\linewidth]{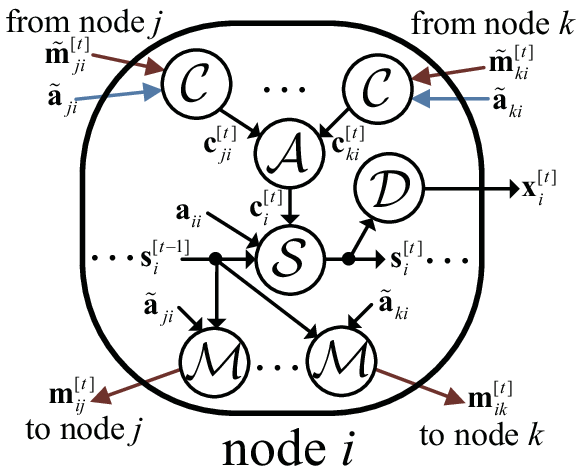}\label{fig:fig2b}
    }
    \subfigure[Message passing structure]{
        \includegraphics[width=.26\linewidth]{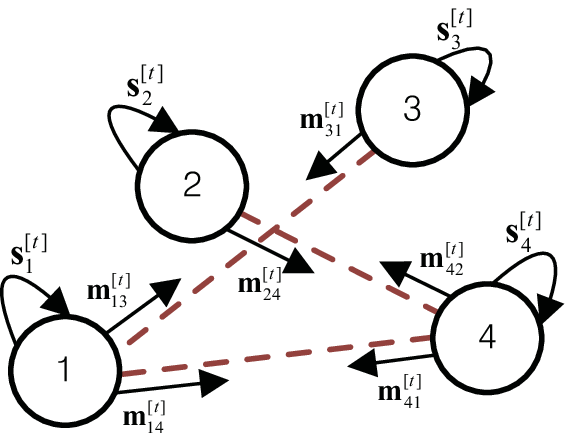}\label{fig:fig2a}
    }
    \subfigure[Graphical interpretation]{
        \includegraphics[width=.3\linewidth]{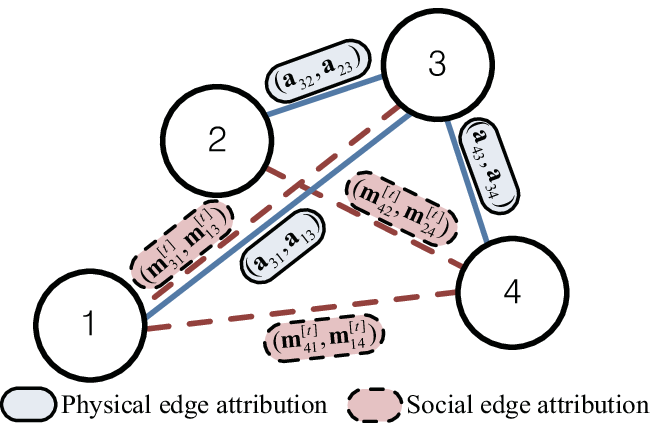}\label{fig:fig2c}
    }
    \caption{Proposed distributed message passing inference.}
    \label{fig:fig2}
\end{figure*}

Fig. \ref{fig:fig2} demonstrates the proposed DMP inference framework that determines the optimal distributed computation rules for (P) over non-complete social graph $\mathcal{G}_{S}$.
Individual DMP nodes calculate their own solutions by passing relevant information iteratively through backhaul links. The DMP inference involves five message computation operations: generation $\mathcal{M}:\mathbb{R}^{S+K_{2}}\rightarrow\mathbb{R}^{M}$, combination $\mathcal{C}:\mathbb{R}^{M+K_{2}}\rightarrow\mathbb{R}^{C}$, aggregation $\mathcal{A}:\mathbb{R}^{C}\rightarrow\mathbb{R}^{C}$, update $\mathcal{S}:\mathbb{R}^{S+C+K_{1}}\rightarrow\mathbb{R}^{S}$, and decision $\mathcal{D}:\mathbb{R}^{S}\rightarrow\mathbb{R}^{X}$. Note that all nodes share these operators to ensure the permutation equivariant property which allows a DL framework to become flexible and robust to network topology change and scaling. Let $\mathcal{N}(i)\triangleq\mathcal{N}_{P}(i)\cup\mathcal{N}_{S}(i)$ be the neighborhood of node $i$ in physical and social domains. A hyperparameter $T$ limits the maximum allowable number of iterations which determines the coordination range of individual nodes. At the $t$-th iteration ($t=1,\ldots,T$), node $i$ handles internal state vector $\mathbf{s}_{i}^{[t-1]}$ to access local knowledge encapsulated at the previous iteration. 
The overall computation rules at the $t$-th iteration are summarized as
\begin{align}
&\textit{Generation: }\!\mathbf{m}_{ij}^{[t]}\!=\!\mathcal{M}(\mathbf{s}_{i}^{[t-1]}\!,\tilde{\mathbf{a}}_{ji}),\!\ j\in\mathcal{N}_{S}(i), \label{eq:mijta}\\
&\textit{Combination: }\mathbf{c}_{ji}^{[t]}=\mathcal{C}(\tilde{\mathbf{m}}_{ji}^{[t]},\tilde{\mathbf{a}}_{ji}),\ j\in\mathcal{N}(i), \label{eq:cjita}\\
&\textit{Aggregation: }\mathbf{c}_{i}^{[t]}=\mathcal{A}(\{\mathbf{c}_{ji}^{[t]}:j\in\mathcal{N}(i)\}),\label{eq:cita}\\
&\textit{Update: }\mathbf{s}_{i}^{[t]}=\mathcal{S}(\mathbf{s}_{i}^{[t-1]},\mathbf{c}_{i}^{[t]},\mathbf{a}_{ii}),\label{eq:sita}\\
&\textit{Decision: }\mathbf{x}_{i}^{[t]}=\mathcal{D}(\mathbf{s}_{i}^{[t]}).\label{eq:xita}
\end{align}
where 
information vectors $\tilde{\mathbf{a}}_{ji}$ and $\tilde{\mathbf{m}}_{ji}^{[t]}$ are defined, respectively, as
\begin{align}
\tilde{\mathbf{a}}_{ji}\label{eq:aji_tilde}
=\begin{cases}
    \mathbf{a}_{ji} &\text{if}~j\in\mathcal{N}_{P}(i),\\
    \mathbf{0}_{K_{2}} &\text{otherwise},
\end{cases}\\
\tilde{\mathbf{m}}_{ji}^{[t]}\label{eq:mjit_tilde}
=\begin{cases}
    \mathbf{m}_{ji}^{[t]} &\text{if}~j\in\mathcal{N}_{S}(i),\\
    \mathbf{0}_{M} &\text{otherwise}.
\end{cases}
\end{align}
The derivation of the above computation rules is presented in Appendix \ref{app:appB}. 
Fig. \ref{fig:fig2b} illustrates the forwardpass computational structure of individual operators.

\subsection{Generation}
At the $t$-th iteration, each node $i$ generates message $\mathbf{m}_{ij}^{[t]}$ dedicated to adjacent node $j\in\mathcal{N}_{S}(i)$ as shown in Fig. \ref{fig:fig2a}. The message generation in \eqref{eq:mijta} encodes $\mathbf{m}_{ij}^{[t]}$ in terms of the previous state $\mathbf{s}_{i}^{[t-1]}$ and the local information that node $i$ senses passively about a physical neighbor $j$, $\mathbf{a}_{ji}$. In case of the lack of the physical neighborhood, the second input of the operator $\mathcal{M}(\cdot)$ becomes a null vector. Such a hybrid input feature informs the physical domain local connectivity $\mathcal{N}_{P}(i)$ to node $i$. The resulting message $\mathbf{m}_{ij}^{[t]}$ is transferred to a social neighbor $j\in\mathcal{N}_{S}(i)$ through a backhaul link. This operation performs active interaction policies such as backhaul signaling and message exchange protocols.
Since all nodes share an identical operator, the message generator $\mathcal{M}(\cdot)$ obtains an access to the global connectivity information $\mathcal{G}_{P}$ during training computation. As a result, the message generation rule in \eqref{eq:mijta} ensures graph-specific calculations according to the network topology. The message propagates to the social neighborhood via backhaul links, and the iterative message propagation allows a multi-hop message exchange between a pair of nodes missing an interconnecting link. This is in essence a crucial mechanism to share the local information globally over the social domain for the DMP inference. 



\subsection{Combination}
The combination operation in \eqref{eq:cjita} consolidates the knowledge about multiplex network configurations and distills essential features for the distributed decision. It puts together all information obtained from social and physical neighborhoods, such as $\mathbf{m}_{ji}^{[t]}$ and $\mathbf{a}_{ji}$. 
The resulting output $\mathbf{c}_{ji}^{[t]}$ of length $C$ can be regarded as an integrated information of physical and social interactions. Since the combination applies to any neighbors in $\mathcal{N}(i)\triangleq\mathcal{N}_{S}(i)\cup\mathcal{N}_{P}(i)$, zero vectors can be fed into either of input variables according to the presence of social and physical neighborhoods. Therefore, the resulting operator can learn the connections in both domains.

\subsection{Aggregation}
The aggregation operation in \eqref{eq:cita} converts the set of incoming messages $\{\mathbf{c}_{ji}^{[t]}:j\in\mathcal{N}(i)\}$ into an aggregate message $\mathbf{c}_{i}^{[t]}$.
Since each node has random interactions with the social neighborhood, the size of incoming message set $\{\mathbf{c}_{ji}^{[t]}:j\in\mathcal{N}(i)\}$ varies with nodes. For the universal inference, the aggregation operator $\mathcal{A}(\cdot)$ can take the input independent of the structure of both domains, which leads to a set operator of incoming message set $\{\mathbf{c}_{ji}^{[t]}:j\in\mathcal{N}(i)\}$ given by
\begin{align}
\mathbf{c}_{i}^{[t]}\!=\!\mathcal{A}(\{\mathbf{c}_{ji}^{[t]}\!:\!j\in\mathcal{N}(i)\})\!=\!\mathcal{A}(\{\mathbf{c}_{\pi(j)i}^{[t]}\!:\!\pi(j)\in\mathcal{N}(i)\}).\label{eq:pia}
\end{align}
Note that the permutation invariance holds with all neighborhood in $j\in\mathcal{N}(i)$.

\subsection{Update and Decision}
The state update operation in \eqref{eq:sita} updates the state $\mathbf{s}_{i}^{[t]}$ in a recursive manner. Node $i$ collects all available knowledge including the previous state $\mathbf{s}_{i}^{[t-1]}$ and the aggregate message $\mathbf{c}_{i}^{[t]}$ associated with the decisions of the social neighborhood, along with the local information $\mathbf{a}_{ii}$ from the physical environment. A new state is determined using the update operator $\mathcal{S}(\cdot)$ with this knowledge collection input. Since the new state $\mathbf{s}_{i}^{[t]}$ contains all information available at node $i$, it suffices to make the final decision about the solution $\mathbf{x}_{i}^{[t]}\in\mathbb{R}^{X}$ only with $\mathbf{s}_{i}^{[t]}$ as in \eqref{eq:xita}.

It can be verified via the node permutation that the solution of the DMP inference achieved by operations in \eqref{eq:mijta}-\eqref{eq:xita} adapts according to the change of the network.
\begin{prop}\label{prop:prop01}
For the DMP inference with a permutation invariant aggregation, the corresponding solution $\mathbf{x}_{i}^{[t]}$ satisfies the permutation equivariance, i.e., $\mathbf{x}_{i}^{[t]}=\mathbf{x}_{\pi(i)}^{\pi[t]},~i\in\mathcal{V}$.
\end{prop}
\begin{IEEEproof}
Since the decision $\mathcal{D}(\cdot)$ in \eqref{eq:xita} accepts only a single input $\mathbf{s}_{i}^{[t]}$, it suffices to show that it takes a common input, i.e., $\mathbf{s}_{i}^{[t]}=\mathbf{s}_{\pi(i)}^{\pi[t]}$. From the permutation invariance, it holds that
\begin{align}
\mathbf{s}^{\pi[t]}_{\pi(i)}&=\mathcal{S}(\mathbf{s}^{\pi[t-1]}_{\pi(i)},\mathbf{c}_{\pi(i)}^{\pi[t]},\mathbf{a}^{\pi}_{\pi(i)\pi(i)})\nonumber\\
    &=\mathcal{S}(\mathbf{s}^{\pi[t-1]}_{\pi(i)}\!,\mathcal{A}(\{\mathbf{c}^{\pi[t]}_{j\pi(i)}\!:\!j\in\mathcal{N}^{\pi}(\pi(i))\}),\!\mathbf{a}^{\pi}_{\pi(i)\pi(i)})\nonumber\\
    &=\mathcal{S}(\mathbf{s}^{[t-1]}_{i},\mathcal{A}(\{\mathbf{c}_{\pi^{-1}(j)i}^{[t]}:\pi^{-1}(j)\in\mathcal{N}(i)\}),\mathbf{a}_{ii})\label{eq:invpi}\nonumber\\
    &=\mathcal{S}(\mathbf{s}^{[t-1]}_{i},\mathcal{A}(\{\mathbf{c}_{ji}^{[t]}:j\in\mathcal{N}(i)\}),\mathbf{a}_{ii})\\
    &=\mathcal{S}(\mathbf{s}^{[t-1]}_{i},\mathbf{c}_{i}^{[t]},\mathbf{a}_{ii})=\mathbf{s}_{i}^{[t]},
\end{align}
where $\pi^{-1}:\mathcal{V}\rightarrow\mathcal{V}$ stands for the inverse of a permutation $\pi$ and \label{eq:invpi} is attained due to the facts of $\mathcal{N}^{\pi}(\pi(i))=\mathcal{N}(i)$ and $z^{\pi}_{ji}=z_{\pi^{-1}(j)\pi^{-1}(i)}$ for a quantity $z_{ji}$. Therefore, the state is obtained accordingly  subject to the graphs change with permutation $\pi$.
\end{IEEEproof}
This statement indicates that the proposed DMP inference can be configured to possess the permutation equivariance in Proposition \ref{prop:prop0}. Thus, its computational structure can be trained to become invariant with social and physical neighborhoods as well as the node population. In particular, node operations $\mathcal{M}(\cdot)$ in \eqref{eq:mijta} and $\mathcal{C}(\cdot)$ in \eqref{eq:cjita} measure the local connectivity information through hybrid input features \eqref{eq:aji_tilde} and \eqref{eq:mjit_tilde}, thereby allowing the state update and the decision to obtain the resulting distributed solution specific to graphs $\mathcal{G}_{P}$ and $\mathcal{G}_{S}$. 

\begingroup
\renewcommand{\baselinestretch}{1.3}
\begin{algorithm}
\caption{DMP Inference Algorithm}
\begin{algorithmic}\label{alg:alg1}
    \STATE Initialize $t=0$ and $\mathbf{s}_{i}^{[0]}$, $\forall i\in\mathcal{V}$.
    \FOR{$t=1,\cdots,T$}
        \STATE {\em Generation:} Each node $i$ generates messages $\mathbf{m}_{ij}^{[t]}=\mathcal{M}(\mathbf{s}_{i}^{[t-1]},\tilde{\mathbf{a}}_{ji})$ for $j\in\mathcal{N}_{S}(i)$, and forwards them through network links.
        \STATE {\em Combination:} Each node $i$ combines the knowledge using $\mathbf{c}_{ji}^{[t]}=\mathcal{C}(\tilde{\mathbf{m}}_{ji}^{[t]},\tilde{\mathbf{a}}_{ji})$ for $j\in\mathcal{N}(i)$.\\
        \STATE {\em Update:} Each node $i$ updates the state using $\mathbf{s}_{i}^{[t]}=\mathcal{S}(\mathbf{s}_{i}^{[t-1]},\mathbf{c}_{i}^{[t]},\mathbf{a}_{ii})$.\\
        \STATE {\em Decision:} Each node $i$ makes the distributed decision with $\mathbf{x}_{i}^{[t]}=\mathcal{D}(s_{i}^{[t]})$.
    \ENDFOR
\end{algorithmic}
\end{algorithm}
\endgroup

Algorithm \ref{alg:alg1} summarizes the computational procedure of the DMP inference. The algorithm begins with an initialization of state variables $\mathbf{s}_{i}^{[0]}$. At each iteration, each node $i$ uses \eqref{eq:mijta} to generate a message $\mathbf{m}_{ij}^{[t]}$ for node $j\in\mathcal{N}_{S}(i)$ and subsequently forwards it to a social neighbor $j \in \mathcal{N}_S(i)$. Upon the message reception, each node conducts the combination, the aggregation, the recursive state update, and the local solution decision. This series of operations is repeated as many times as the predefined iteration number $T$.





\section{Message-Passing Neural Networks}\label{sec:sec5}
In this section, we develop a model-driven DMPNN framework that learns the DMP inference in Algorithm \ref{alg:alg1} for the network utility optimization in (P). The DMPNN conducts node operations $\mathcal{M}(\cdot)$, $\mathcal{C}(\cdot)$, $\mathcal{A}(\cdot)$, $\mathcal{S}(\cdot)$, and $\mathcal{D}(\cdot)$ in \eqref{eq:mijta}-\eqref{eq:xita} in terms of DNNs. The node operations of message generation, combination, aggregation, and decision functions employ FNN structures, as opposed to the state update operation realized in an RNN. Let $\text{FNN}_{L}(\mathbf{z};\mathbf{\theta})$ denote an $L$-layer fully-connected FNN with input $\mathbf{z}$ and parameter set $\mathbf{\theta}$. 
The FNNs implementing node operations $\mathcal{M}(\cdot)$, $\mathcal{C}(\cdot)$, and $\mathcal{D}(\cdot)$
in \eqref{eq:mijta}, \eqref{eq:cjita}, and \eqref{eq:xita} are denoted by $\text{FNN}_{L_{M}}(\cdot;\mathbf{\theta}_{M})$, $\text{FNN}_{L_{C}}(\cdot;\mathbf{\theta}_{C})$, and $\text{FNN}_{L_{D}}(\cdot;\mathbf{\theta}_{D})$, respectively, and the resulting node operations are represented as
\begin{align}
\mathbf{m}_{ij}^{[t]}&=\mathcal{M}(\mathbf{s}_{i}^{[t-1]},\tilde{\mathbf{a}}_{ji})=\text{FNN}_{L_{M}}(\mathbf{s}_{i}^{[t-1]},\tilde{\mathbf{a}}_{ji};\mathbf{\theta}_{M}),\label{eq:fnnm}\\
\mathbf{c}_{ji}^{[t]}&=\mathcal{C}(\tilde{\mathbf{m}}_{ji}^{[t]},\tilde{\mathbf{a}}_{ji})=\text{FNN}_{L_{C}}(\tilde{\mathbf{m}}_{ji}^{[t]},\tilde{\mathbf{a}}_{ji};\mathbf{\theta}_{C}),\label{eq:fnnc}\\
\mathbf{x}_{i}^{[t]}&=\mathcal{D}(\mathbf{s}_{i}^{[t]})=\text{FNN}_{L_{D}}(\mathbf{s}_{i}^{[t]};\mathbf{\theta}_{D}).\label{eq:fnnd}
\end{align}
Note that the output layer of $\text{FNN}_{L_{D}}(\cdot;\mathbf{\theta}_{D})$ guarantees the membership of the solution $\mathbf{x}_{i}^{[t]}$ in a solution space $\mathcal{X}$, i.e., $\mathbf{x}_{i}^{[t]}\in\mathcal{X}$, which is realized by a projection activation \cite{HLee:19b}. The following statement assesses DNN-based node operations from the universal approximation theorem \cite{KHornik:89}.
\begin{prop}\label{prop:prop1}
Let $\mathcal{U}(\mathbf{z})$ be a continuous vector function defined over a bounded region $\mathcal{Z}\subset\mathbb{R}^{Z}$. For any $\varepsilon>0$, there exists an $\text{FNN}_{L}(\mathbf{z};\mathbf{\theta})$ with finite $L$ and sigmoid activations such that
\begin{align}
\sup_{\mathbf{z}\in\mathcal{Z}}\|\mathcal{U}(\mathbf{z})-\text{FNN}_{L}(\mathbf{z};\mathbf{\theta})\|\leq\varepsilon.\label{eq:prop1}
\end{align}
\end{prop}
\begin{IEEEproof}
This is an extension of the universal approximation theorem for a scalar-valued target mapping \cite{KHornik:89}. The result for a general vector-valued mapping case in \eqref{eq:prop1} is established by constructing $\text{FNN}_{L}(\mathbf{z};\mathbf{\theta})$ in an array of multiple scalar-valued FNNs, each approximating an individual element of $\mathcal{U}(\mathbf{z})$. Mathematical rigor to complete the statement about the existence is omitted since it proceeds similarly as in \cite{HSun:18}.
\end{IEEEproof}
The characterization of a discrete-valued function by an FNN with rectified linear unit (ReLU) activations has recently been realized \cite{ZLu:17}. This speculates that the FNN approaches in \eqref{eq:fnnm}-\eqref{eq:fnnd} ensure to conduct the corresponding node operations regardless of analytical properties.

The state update operator $\mathcal{S}(\cdot)$ in \eqref{eq:sita} resorts to an RNN structure with a single input of the previous state $\mathbf{s}_{i}^{[t-1]}$. 
A gated recurrent unit (GRU) \cite{KCho:14} is used in that it addresses a long-term dependency of the vanilla RNN with reduced complexity as compared to a long-short term memory technique \cite{IGoodfellow:16}.
The state update operation is represented in a GRU as
\begin{align}
\mathbf{s}_{i}^{[t]}=\mathcal{S}(\mathbf{s}_{i}^{[t-1]},\mathbf{c}_{i}^{[t]},\mathbf{a}_{ii})=\text{GRU}(\mathbf{s}_{i}^{[t-1]},\mathbf{c}_{i}^{[t]},\mathbf{a}_{ii};\mathbf{\theta}_{S}).\label{eq:grus}
\end{align}

The aggregation operator $\mathcal{A}(\cdot)$ in \eqref{eq:cita} has a different DNN structure from other operators so that the forwardpass computation is independent of node population $N$ and interaction models $(\mathcal{G}_{P},\mathcal{G}_{S})$. As discussed, the aggregation operation at node $i$ becomes a function of  incoming message set $\{\mathbf{c}_{ji}^{[t]}:j\in\mathcal{N}(i)\}$, yielding an output that does not vary with the order of input messages, i.e., the condition in \eqref{eq:pia} holds. However, a conventional DNN structure fails to capture this because input dimension is fixed and, when the social domain changes, a new DNN structure is trained over with new configurations of node population and backhaul topology.

To realize such an invariance in the aggregation operator $\mathcal{A}(\cdot)$, we consider a notion of a {\em measure} on a message set $\{\mathbf{c}_{ji}^{[t]}:j\in\mathcal{N}(i)\}$ from real analysis. The measure evaluates a score of an input set by assigning a specific number, and its essential property is the additivity. Let $\mathcal{Z}_{q}\subset\mathcal{Z}$ $(q=1,\cdots,Q)$ be disjoint subsets of a set $\mathcal{Z}$. Then, the additivity of measure $\mathcal{A}(\cdot)$ is ensured by $\mathcal{A}(\bigcup_{q=1}^{Q}\mathcal{Z}_{q})=\sum_{q=1}^{Q}\mathcal{A}(\mathcal{Z}_{q})$ \cite{Rudin:87}, i.e., the measure of the union of disjoint sets is identical to the sum of their individual measures. Substituting this into \eqref{eq:cita} leads to
\begin{align}
\mathbf{c}_{i}^{[t]}=\mathcal{A}(\{\mathbf{c}_{ji}^{[t]}:j\in\mathcal{N}(i)\})
                =\sum_{j\in\mathcal{N}(i)}\mathcal{A}(\{\mathbf{c}_{ji}^{[t]}\}),\label{eq:additivity}
\end{align}
where singleton sets $\{\mathbf{c}_{ji}^{[t]}\}$ $(j\in\mathcal{N}(i))$ become disjoint subsets of the universal message set $\{\mathbf{c}_{ji}^{[t]}:j\in\mathcal{N}(i)\}$. As desired, \eqref{eq:additivity} varies with neither the ordering of incoming messages nor the overall populations of physical and social neighborhoods. Since the aggregation operator $\mathcal{A}(\cdot)$ takes $\mathbf{c}_{ji}^{[t]}$ as a sole input, its computation is readily modeled by a typical FNN denoted by $\text{FNN}_{L_{A}}(\cdot;\mathbf{\theta}_{A})$ and expressed as $\mathbf{c}_{i}^{[t]}=\sum_{j\in\mathcal{N}(i)}\text{FNN}_{L_{A}}(\mathbf{c}_{ji}^{[t]};\mathbf{\theta}_{A})$.
The computation of $\text{FNN}_{L_{A}}(\cdot;\mathbf{\theta}_{A})$ can be combined with the combination operation in \eqref{eq:fnnc}. This reduces the aggregation strategy to\footnote{The aggregation rule is applicable to local observation vector $\mathbf{a}_{i}$ to adapt universal input dimension.}
\begin{align}
\mathbf{c}_{i}^{[t]}=\mathcal{A}(\{\mathbf{c}_{ji}^{[t]}:j\in\mathcal{N}(i)\})=\sum_{j\in\mathcal{N}(i)}\mathbf{c}_{ji}^{[t]},\label{eq:sumc}
\end{align}
and the state update operation in \eqref{eq:grus} is refined as
\begin{align}
\mathbf{s}_{i}^{[t]}=\mathcal{S}(\mathbf{s}_{i}^{[t-1]},\mathbf{c}_{i}^{[t]},\mathbf{a}_{ii})
    =\text{GRU}\bigg(\mathbf{s}_{i}^{[t-1]},\sum_{j\in\mathcal{N}(i)}\mathbf{c}_{ji}^{[t]},\mathbf{a}_{ii};\mathbf{\theta}_{S}\bigg).\label{eq:grus2}
\end{align}
Therefore, the forwardpass computation and the DMP inference in Algorithm \ref{alg:alg1} are realized in a DNN structure that combines \eqref{eq:fnnm}-\eqref{eq:fnnd} and \eqref{eq:grus2}. The forwardpass computation iterates up to the maximum steps of $T$ with an identical DNN parameter arragement.

Note that the permutation-invariant aggregation operation successfully realizes the measure of a set, which is the key enabler for the universality of the DMPNN. The following statement consolidates the effectiveness of the proposed sum aggregation strategy in \eqref{eq:sumc}. It suggests an alternative formulation of the aggregation operation that satisfies the permutation invariance property and indeed constructs a generic form of the set function decomposed in \eqref{eq:prop2}.
\begin{prop}\label{prop:prop2}
A set function $\mathcal{U}(\{\mathbf{z}_{1},\cdots,\mathbf{z}_{Q}\})$ defined over a set of discrete variables $\mathbf{z}_{q}\in\mathcal{Z}$ with a finite alphabet 
is permutation invariant if and only if it can be decomposed into
\begin{align}
\mathcal{U}(\{\mathbf{z}_{1},\cdots,\mathbf{z}_{Q}\})=\rho\Bigg(\sum_{q=1}^{Q}\phi(\mathbf{z}_{q})\Bigg)\label{eq:prop2}
\end{align}
for some mappings $\rho(\cdot)$ and $\phi(\cdot)$.
\end{prop}
\begin{IEEEproof}
The sufficiency of the statement is straightforward: The right hand side of \eqref{eq:prop2} is invariant with the permutation of set elements. Hence, the transformation $\rho(\sum_{q=1}^{Q}\phi(\mathbf{z}_{q}))$ naturally secures the permutation invariant property. The necessity of the existence of a decomposing pair of $\rho(\cdot)$ and $\phi(\cdot)$ begins with indexing input elements. Since the input is indeed a set of discrete elements, it can be represented in a unique number. Thus, there exists a mapping $\phi(\cdot)$ evaluating the value of each digit, and the resulting value corresponds to the sum aggregation $\sum_{q=1}^{Q}\phi(\mathbf{z}_{q})$. Furthermore, an outer mapping $\rho(\cdot)$ exists to calculate the output of the set function. Therefore, there always exist $\rho(\cdot)$ and $\phi(\cdot)$ that can characterize the set function.
\end{IEEEproof}

This property can be generalized to a continuous input 
along with its universal approximation property for both discrete and continuous input cases \cite[Theorem 9]{MZaheer:17}. These properties suggest that aggregation operator $\mathcal{A}(\cdot)$ is of a formulation as in \eqref{eq:prop2}, and two component functions $\phi(\cdot)$ and $\rho(\cdot)$ are configured with high accuracy via DNNs of $\text{FNN}_{L_{\rho}}(\cdot;\mathbf{\theta}_{\rho})$ and $\text{FNN}_{L_{\phi}}(\cdot;\mathbf{\theta}_{\phi})$, respectively. 
Subsequently, the aggregation operator $\mathcal{A}(\cdot)$ is expressed as
\begin{align} \label{opci}
\mathbf{c}_{i}^{[t]}=&\mathcal{A}(\{\mathbf{c}_{ji}^{[t]}: j\in\mathcal{N}(i)\})\nonumber\\ =&\text{FNN}_{L_{\rho}}\bigg(\sum_{j\in\mathcal{N}(i)}\text{FNN}_{L_{\phi}}(\mathbf{c}_{ji}^{[t]};\mathbf{\theta}_{\rho});\mathbf{\theta}_{\phi}\bigg).
\end{align}
Note here that $\text{FNN}_{L_{\phi}}(\cdot;\mathbf{\theta}_{\phi})$ can be integrated with the combination operation in \eqref{eq:fnnc}. Furthermore, since the message aggregate $\mathbf{c}_{i}^{[t]}$ is fed into the GRU in \eqref{eq:grus}, the forwardpass computation for $\text{FNN}_{L_{\rho}}(\cdot;\mathbf{\theta}_{\rho})$ can be incorporated into the GRU. This again leads to the sum aggregation mechanism captured by the measure theory-inspired design in \eqref{eq:sumc}. Therefore, the permutation-invariant set aggregation enables the DMPNN to adapt to random graphs $\mathcal{G}_{P}$ and $\mathcal{G}_{S}$.

\subsection{Training and Implementation}
We present a training strategy and a distributed implementation of the DMPNN. 
Let $\mathbf{\Theta}\triangleq\{\mathbf{\theta}_{M},\mathbf{\theta}_{C},\mathbf{\theta}_{S},\mathbf{\theta}_{D}\}$ be the hyperparameter set of the DMPNN. To capture the iterative nature, the objective function $\mathcal{F}(\mathbf{\Theta})$ is defined to assess a tentative solution $\mathbf{x}^{[t]}\triangleq\{\mathbf{x}_{i}^{[t]}: i\in\mathcal{V}\}$ obtained at the $t$-th iteration $(t=1,\cdots,T)$. To this end, we introduce an increasing weight $\sqrt{t}$ to the utility $f(\mathbf{a},\mathbf{x}^{[t]})$ in the original formulation (P) so that the DMPNN approaches an efficient solution of the network utility maximization problem (P) as $t$ grows. The corresponding objective function is given by a weighted sum utility formulation as
\begin{align}
\mathcal{F}(\mathbf{\Theta})=\mathbb{E}_{\mathbf{a},\mathcal{G}_{S}}\bigg[\sum_{t=1}^{T}\sqrt{t}f(\mathbf{a},\mathbf{x}^{[t]})\bigg].\label{eq:obj}
\end{align}
Here, the weighted sum utility in \eqref{eq:obj} is used widely for handling iterative and recursive learning tasks \cite{MZaheer:18}. 
Furthermore, it is also known to overcome vanishing gradient issues encountered during the training of very deep network structure \cite{CSzegedy:15}. The training algorithm proceeds with mini-batch stochastic gradient descent (SGD) as
\begin{align}
\mathbf{\Theta}\leftarrow\mathbf{\Theta}+\eta\mathbb{E}_{\mathcal{B}}\Bigg[\sum_{t=1}^{T}\sqrt{t}\nabla_{\mathbf{\Theta}} f(\mathbf{a},\mathbf{x}^{[t]})\Bigg],\label{eq:sgd}
\end{align}
where $\mathcal{B}$ is a mini-batch containing $B=|\mathcal{B}|$ independently generated training samples $(\mathbf{a},N,\mathcal{G}_{S})$. 

The DMPNN is trained in a fully unsupervised manner since no labeled data point regarding the optimal solution of (P) is necessary in training computations.
However, full and global knowledge about physical and social interactions is necessary. In practice, the global topology information is collected from individual nodes and stored beforehand at central agents such as a DL cloud. Alternatively, it can be sampled at random with the probability distribution of local information $\mathbf{a}$, such as Rayleigh fading channel gains. Social connection $\mathcal{G}_{S}$ and node population $N$ can also be randomly generated by the cloud for the mini-batch collection. Thus, the cloud can conduct the SGD update in \eqref{eq:sgd} in an offline manner ahead of the online forwardpass computation. To facilitate its implementation and enhance parallel computing capability of existing DL libraries, out of a complete two-layer multiplex graph with $N_{\max}$ node population, $N_{\max}-N$ nodes are randomly decimated and, subsequently, edges of physical and social domains are also independently removed according to their connection probabilities.
During training the DMPNN, in particular, the message generation and the combination can access to full graph connection topologies shared at all nodes. with numerous instances of $\mathcal{G}_{P}$ and $\mathcal{G}_{S}$, they learn cooperation and decision policies for arbitrary interaction models.

The trained parameter set $\mathbf{\Theta}$ is stored in memory units of individual nodes. Note that, in this offline training process, nodes are enabled to cooperate with a possible range of neighborhood population. Thus, their computation time and efforts in message processing are unaffected by network topology change and scaling, since the neighborhood population and the incoming message set remain in a manageable level. Individual nodes yield a real-time inference of the DMPNN trained to run Algorithm \ref{alg:alg1}. This online DMP inference uses only local information such as $\mathbf{a}_{i}$, $\mathcal{N}_{P}(i)$, and $\mathcal{N}_{S}(i)$ for a distributed solution of (P). Since the forwardpass computation does not vary with $N$, $\mathcal{G}_{P}$, and $\mathcal{G}_{S}$, the online operation is carried out locally at each node and is extended universally to arbitrary network configurations.

\subsection{Relationship with Graph Neural Networks}\label{sec:sec5b}

By taking physical and social graphs as inputs to produce a network solution $\mathbf{x}$, the proposed framework can be viewed as a class of generalized GNNs, which is considered for graph based tasks in various applications \cite{CQi:17,MZaheer:18,JGilmer:17}. Several GNN variants address wireless communication applications, such as nonconvex power control \cite{YShen:19,MEisen:19,AChowdhury:21} and link scheduling \cite{MLee:21}. Although feedforward-type DNN structures improve the scalability and the expressive power of graph based tasks, these works lack decentralized operation strategies for practical deployment. Furthermore, interaction mechanisms focus only on network configurations of identical physical and social domain topology. The underlying GNN models are expressed as
\begin{align}
\mathbf{c}_{i}^{[t]}=&\mathcal{A}(\{\mathbf{s}_{j}^{[t-1]}:j\in\mathcal{N}(i)\}),\ \forall i\in\mathcal{V},\\
\mathbf{s}_{i}^{[t]}=&\mathcal{S}(\mathbf{s}_{i}^{[t-1]},\mathbf{c}_{i}^{[t]}),\ \forall i\in\mathcal{V}.\label{eq:gnnc}
\end{align}
Note that the state $\mathbf{s}_{i}^{[t]}$ in \eqref{eq:gnnc} acts as a message since it is distributed from node $i$ to the neighborhood $\mathcal{N}(i)$. However, the solution $\mathbf{x}$ is centrally determined at the final iteration $t=T$~as
\begin{align}
\mathbf{x}=\{\mathbf{x}_{i}:i\in\mathcal{V}\}=\mathcal{D}(\{\mathbf{s}_{i}^{[T]}:i\in\mathcal{V}\}).\label{eq:gnnd}
\end{align}
The forwardpass computations in \eqref{eq:gnnc} and \eqref{eq:gnnd} do not access physical attributes such as $\{\mathbf{a}_{ji}\}$. Since the wireless network performance is essentially affected by the physical domain topology, conventional GNNs fail to address network management problems (P1) and (P2) for their dependence on  centralized operations of information collection, solution identification, and distribution.

Although several works have addressed physical attributes as additional input feature \cite{YShen:19,JGilmer:17}, those approaches lack the consideration on the distributed implementation of GNNs as well as random networking configurations with varying $N$ and $\mathcal{G}_{S}$. Interference graph convolutional network (IGCN) intended for the sum-rate maximization (P1) is formulated in \cite{YShen:19} as
\begin{align}
\mathbf{c}_{ij}^{[t]}&=\mathcal{C}(\mathbf{s}_{i}^{[t-1]},\mathbf{a}_{ji},\mathbf{a}_{ij},\mathbf{a}_{ii}),\ i\in\mathcal{V},j\in\mathcal{N}(i),\label{eq:igcnc}\\
\mathbf{c}_{i}^{[t]}&=\bigg(\sum_{j\in\mathcal{N}(i)}\mathbf{c}_{ji}^{[t]},\max_{j\in\mathcal{N}(i)}\mathbf{c}_{ji}^{[t]}\bigg),\ i\in\mathcal{V},\label{eq:igcna}\\
\mathbf{s}_{i}^{[t]}&=\mathcal{S}(\mathbf{s}_{i}^{[t-1]},\mathbf{c}_{i}^{[t]},\mathbf{a}_{ii}),\ i\in\mathcal{V},\label{eq:igcns}
\end{align}
where the aggregation in \eqref{eq:igcna} is the concatenation of sum-pooling and max-pooling of incoming messages dedicated to each node $i$, and $\mathcal{C}(\cdot)$ and $\mathcal{S}(\cdot)$ are implemented with FNNs.

The MPNN \cite{JGilmer:17} learns chemical properties with an inference structure expressed as
\begin{align}
\mathbf{c}_{ij}^{[t]}&=\mathcal{C}(\mathbf{s}_{i}^{[t-1]},\mathbf{s}_{j}^{[t-1]},\mathbf{a}_{ji},\mathbf{a}_{ij}),\ i\in\mathcal{V},j\in\mathcal{N}(i),\label{eq:mpnnc}\\
\mathbf{s}_{i}^{[t]}&=\mathcal{S}\bigg(\mathbf{s}_{i}^{[t-1]},\sum_{j\in\mathcal{N}(i)}\mathbf{c}_{ji}^{[t]}\bigg),\ i\in\mathcal{V},\label{eq:mpnns}
\end{align}
where $\mathcal{S}(\cdot)$ uses the sum-pooling aggregation.
Note here that combination operations $\mathcal{C}(\cdot)$ in \eqref{eq:igcnc} and \eqref{eq:mpnnc} indeed resort to centralized coordinations since each node $i$ necessarily knows internal state $\mathbf{s}_{j}^{[t-1]}$ and local observation $\mathbf{a}_{ij}$ of its neighbor $j$. Under the assumption of uniform topology of physical and social domains, raw local information can be directly exchanged among nodes. State updates $\mathcal{S}(\cdot)$ entail reciprocal exchanges of $\mathbf{c}_{ij}^{[t]}$ for each pair of connected nodes $i$ and $j$, incurring additional signaling overheads. In addition, the MPNN makes centralized decisions of all solutions in \eqref{eq:gnnd} with the collection of all node state information. 
On the other hand, the DMPNN is exempt from these restrictions and can be viewed as a generalized extension of existing techniques for universal distributed tasks. The beauty of the proposed framework stems from the decoupling property of the message computations such as \eqref{eq:igcnc} and \eqref{eq:mpnnc} into the message generation in \eqref{eq:fnnm} and the combination in \eqref{eq:fnnc}.
This allows local message computations and distributed optimizations in multiplex networks with physical and social domains.

Scalable GNN architectures are studied for tackling (P1) \cite{MEisen:19,AChowdhury:21}. Interference from transmitters to receivers is modeled by a bipartite graph having two disjoint vertex sets, i.e., sets of transmitter nodes and receiver nodes. The random edge GNN (REGNN) is developed for addressing randomly varying physical domain $\mathcal{G}_{P}$ \cite{MEisen:19}. 
The forwardpass computation rule for the collection of scalar states $s_{i}^{[t]}$  $\mathbf{s}^{[t]}=\{s_{i}^{[t]}:i\in\mathcal{V}\}$ is given at the $t$-th iteration by
\begin{align}
\mathbf{s}^{[t]}=\sigma\bigg(\sum_{q=0}^{Q}w_{q}^{[t]}\mathbf{A}^{q}\mathbf{s}^{[t-1]}\bigg),\label{eq:regnn}
\end{align}
where $\sigma(\cdot)$ denotes an activation, $w_{q}^{[t]}$ is a trainable weight of the $q$-th filter, and the matrix $\mathbf{A}$ describes the physical relationships between transmitters and receivers. Its $(i,j)$-element is set to a scalar local observation $a_{ij}$ if $(i,j)\in\mathcal{E}_{P}$, i.e., when transmitter $i$ interferes receiver $j$, and zero, otherwise. The solution is directly obtained from state variables $\mathbf{x}=\mathbf{s}^{[T]}$. Handling matrix $\mathbf{A}^{q}$ involves the collection of global physical domain graph topology and centralized matrix computations. Otherwise, it would be realized in a distributed manner if both transmitters and receivers participate in a cooperative optimization. Since $Q$ downlink-uplink communication rounds are carried out at a single iteration, the total of $QT$ communication rounds is necessary for the final solution. This might be burden for battery-powered receiver devices. On the contrary, only transmitters are responsible for the computations in the proposed framework. This can be achieved by the proposed DMP inference that extends the message passing architecture of GNNs to the multiplex network setup with individual physical and social graphs.

Graph embedding for binary-link scheduling \cite{MLee:21} is considered over a single coupled graph jointly representing physical and social interactions. Node states are graph embeddings obtained using similar update rules with \eqref{eq:regnn}. Subsequently, the centralized solution is realized by the collection of all node embeddings.

\section{Numerical Results}\label{sec:sec6}
We test the performance of the DMPNN for tackling two power control problems of (P1) and (P2). The node population corresponding to the number of the transmitter-receiver pairs in the wireless network is chosen at random within the range of $N \in [3,10]$. The maximum transmit power is set to $P=10$, and the channel gain $a_{ji}$ is an exponential random variable with unit mean. Unless stated otherwise, physical domain $\mathcal{G}_{P}$ has a graph of complete connections among all node pairs, i.e., $\mathcal{E}_{P}=\{(i,j):j\in\mathcal{V}\backslash \{i\}\}$. On the other hand, we consider an Erdos-Renyi graph model $(N,p)$ for social domain $\mathcal{G}_{S}$, i.e., each edge $(i,j)\in\mathcal{G}_{S}$ is independently connected with probability $p$. For a performance evaluation, the value of $p$ changes between train and test processes. The corresponding edge probabilities are denoted by $p_{\text{train}}$ and $p_{\text{test}}$, respectively. The dimensions of the message aggregation $\mathbf{c}_{ji}^{[t]}$ and the state $\mathbf{s}_{i}^{[t]}$ are both set to $C=S=50$. Unless stated otherwise, the dimension of message $\mathbf{m}_{ji}^{[t]}$ is set to  $M=10$. The state update operation in \eqref{eq:grus} is implemented in a single layer GRU with the hidden layer dimension $S=50$. The state $\mathbf{s}_{i}^{[0]}$ is initialized as a vector of zero-mean unit-variance Gaussian random variables. The ReLU activation defined as $\text{ReLU}(z)=\max\{0,z\}$ is adopted for all layers of FNNs in \eqref{eq:fnnm}-\eqref{eq:fnnd}, while the output layer of the decision FNN in \eqref{eq:fnnd} is a sigmoid activation $\sigma(z)=\frac{P}{1+e^{-z}}$ for the feasibility of nonnegative power control solutions.

The training algorithm is implemented with Tensorflow. The Adam algorithm \cite{Kingma:15} is used with the learning rate $\eta=0.0001$ and the mini-batch size $B=1000$. The DMPNN is trained during $1000$ training epoches, each consisting of $50$ mini-batches.
The performance of the DMPNN is examined using the validation data set of $10^{4}$ samples.

\subsection{Sum Rate Maximization}
\begin{figure}
\centering
\includegraphics[width=.5\linewidth]{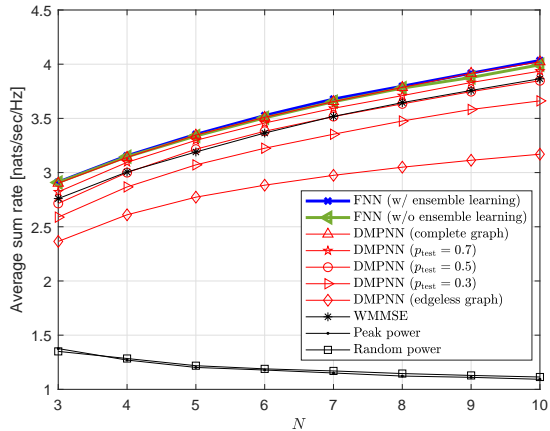}
\caption{Average sum rate as a function of $N$ with $p_{\text{train}}=0.7$.}
\label{simul:fig1}
\end{figure}

We first test the performance for maximum sum-rate problem (P1). Three-layer FNNs are employed for \eqref{eq:fnnm}-\eqref{eq:fnnd} with the hidden layer dimension $100$ and the maximum iteration number $T=20$. Fig. \ref{simul:fig1} depicts the average sum rate performance of the DMPNN trained with $p_{\text{train}}=0.7$ with respect to $p_{\text{test}}$ and $N$. Complete and edgeless graphs are characterized by Erdos-Renyi random graphs with $p_{\text{test}}=1$ and $p_{\text{test}}=0$, respectively. A practical backhaul setup is modeled with $p_{\text{test}}\geq0.5$, since a random graph generation with $p_{\text{test}}=0.5$ leads to a uniform collection of $2^{{N}\choose{2}}$ possible graphs. As benchmarks, the following baseline techniques are compared.
\begin{itemize}
\item {\em FNN:} A typical FNN is trained with ten layers of the hidden-layer dimension $150$. The corresponding system has a similar number of parameters with the DMPNN.
\item {\em WMMSE:} A local optimum is obtained from the WMMSE algorithm \cite{QShi:11}.
\item {\em Peak power:} The transmit power is set to the maximum $x_{i}=P$, $\forall i\in\mathcal{V}$.
\item {\em Random power:} The transmit power is chosen at random over $[0,P]$.
\end{itemize}
The FNN baseline provides the state-of-the-art performance of unsupervised learning methods \cite{WLee:18a,HLee:19b}. Eight different versions of FNNs, each dedicated to one of $N\in[3,10]$, are constructed to operate in a centralized manner. The overall performance can be further improved via an ensemble training technique \cite{FLiang:20}, where five FNNs with different initializations are individually trained, and the best one is chosen out of them. A distributed implementation of the WMMSE algorithm is plausible \cite{QShi:11}, although complete social graph $\mathcal{G}_{S}$ is generally required so that each transmitter pair is interconnected. In comparison with these baselines, the DMPNN adapts to any backhaul configuration with arbitrary value of $N$. Fig. \ref{simul:fig1} show that the distributed DMPNN implementation exhibits the performance similar to the FNN implementation, when tested over the complete social graph. When the DMPNN is tested at $p_{\text{test}}=0.5$ corresponding to a quite challenging condition with traditional optimization techniques, it obtains locally optimum performance resulting from the WMMSE. This verifies that the DMPNN can effectively approximate the DMP inference that involves iterative exchanges of messages over arbitrary social and physical domains. One can see that the DMPNN, when tested in an edgeless setup of no node interaction, still shows a good performance that outperforms naive peak power and random power allocation schemes. In training with Erdos-Renyi graphs, the DMPNN observes numerous instances showing average behaviors of nodes under random link connections including non-complete graphs. This knowledge produces an efficient distributed strategy for the DMPNN even in the connectionless case.

\begin{figure}
\centering
\includegraphics[width=.5\linewidth]{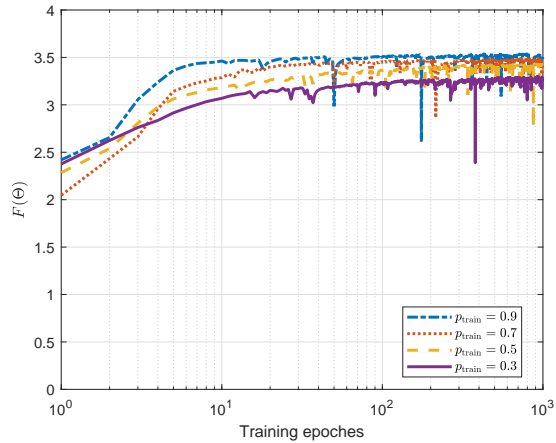}
\caption{Convergence behavior of training process for various $p_{\text{train}}$.}
\label{simul:fig4}
\end{figure}

Fig. \ref{simul:fig4} validates convergence behaviors during the training computations by demonstrating trajectories of the objective function $\mathcal{F}(\mathbf{\Theta})$ in \eqref{eq:obj} for different values of $p_{\text{train}}$. The proposed weighted sum utility in \eqref{eq:obj} converges within $100$ training epoches consistently over random configurations. The number of node connections scales up as $p_{\text{train}}$ increases, thus rising the volume of the shared information over the network. The resulting training objective grows for increasing $p_{\text{train}}$. On the other hands, small $p_{\text{train}}$ degrades the validation performance since the optimization over less-connected backhaul configurations is more challenging.

\begingroup
\renewcommand{\baselinestretch}{1.3}
\renewcommand{\arraystretch}{0.8}
\begin{table*}[!htp]
\centering
\caption{Impact of hidden layer dimension}
\begin{tabular}{c|c|c|c||c|c|c||c|c|c|}
\cline{2-10}
                          & \multicolumn{3}{c||}{$p_{\text{test}}=0.5$}  & \multicolumn{3}{c||}{$p_{\text{test}}=0.7$}  & \multicolumn{3}{c|}{Complete graph}  \\ \hline
\multicolumn{1}{|c|}{$N$} & $(50,50)$ & $(100,50)$ & $(100,100)$ & $(50,50)$ & $(100,50)$ & $(100,100)$ & $(50,50)$ & $(100,50)$ & $(100,100)$ \\ \hline
\multicolumn{1}{|c|}{3}   & 2.733 & 2.754 & \textbf{2.796}          & 2.845 & 2.836 & \textbf{2.857}          & 2.901 & \textbf{2.903} & 2.903       \\ \hline
\multicolumn{1}{|c|}{5}   & 3.231 & 3.218 & \textbf{3.245}          & 3.299 & 3.296 & \textbf{3.301}          & 3.339 & 3.343 & \textbf{3.344}       \\ \hline
\multicolumn{1}{|c|}{7}   & \textbf{3.521} & 3.516 & 3.509          & 3.585 & 3.589 & \textbf{3.590}          & 3.638 & 3.650 & \textbf{3.660}       \\ \hline
\multicolumn{1}{|c|}{9}   & 3.743 & \textbf{3.751} & 3.745          & 3.813 & 3.824 & \textbf{3.828}          & 3.889 & \textbf{3.911} & 3.909       \\ \hline
\end{tabular}
\label{table:table1}
\end{table*}
\endgroup

\begingroup
\renewcommand{\baselinestretch}{1.3}
\renewcommand{\arraystretch}{0.8}
\begin{table*}[htp!]
\centering
\caption{Impact of message dimension $M$}
\subtable[Average sum rate with $p_{\text{train}}=0.6$]{
\centering
\begin{tabular}{|c|c||c||c|c|c|c||c|c|c|c||c|c|c|c|}
\hline
\multirow{3}{*}{$N$} & \multirow{3}{*}{\!\!FNN\!\!} & \multirow{3}{*}{\!\!\!WMMSE\!\!\!} & \multicolumn{12}{c|}{DMPNN ($p_{\text{train}}=0.6$)}                                                                                                                                                                                                                                                                                                                       \\ \cline{4-15}
                     &                      &                        & \multicolumn{4}{c||}{$p_{\text{test}}=0.5$}                                                                                                   & \multicolumn{4}{c||}{$p_{\text{test}}=0.7$}                                                                          & \multicolumn{4}{c|}{Complete graph}                                                                          \\ \cline{4-15}
                     &                      &                        & $\!\!M=3\!\!$                           & $\!\!M=5\!\!$                           & $\!\!M=10\!\!$                          & $\!\!M=15\!\!$                          & $\!\!M=3\!\!$                           & $\!\!M=5\!\!$                           & $\!\!M=10\!\!$                          & $\!\!M=15\!\!$ & $\!\!M=3\!\!$                           & $\!\!M=5\!\!$                           & $\!\!M=10\!\!$                          & $\!\!M=15\!\!$ \\ \hline
3 & 2.909 & 2.760          & 2.735 & 2.754 & 2.754 & \textbf{2.788}         & 2.828 & 2.840 & 2.836 & \textbf{2.847}       & 2.901 & \textbf{2.904} & 2.903 & 2.896  \\ \hline
5 & 3.338 & 3.190          & 3.217 & 3.233 & 3.218 & \textbf{3.236}         & 3.294 & \textbf{3.303} & 3.296 & 3.299       & 3.337 & \textbf{3.345} & 3.343 & 3.341  \\ \hline
7 & 3.654 & 3.518          & 3.508 & \textbf{3.522} & 3.516 & \textbf{3.522}         & 3.587 & \textbf{3.590} & 3.589 & \textbf{3.590}       & 3.641 & 3.648 & \textbf{3.650} & 3.647  \\ \hline
9 & 3.877 & 3.758          & 3.739 & 3.743 & 3.751 & \textbf{3.752}          & 3.821 & 3.820 & \textbf{3.824} & 3.822       & 3.898 & 3.898 & \textbf{3.911} & 3.901  \\ \hline
\end{tabular}\label{table:table2a}
}
\subtable[Average sum rate with $p_{\text{train}}=0.8$]{
\centering
\begin{tabular}{|c|c||c||c|c|c|c||c|c|c|c||c|c|c|c|}
\hline
\multirow{3}{*}{$N$} & \multirow{3}{*}{\!\!FNN\!\!} & \multirow{3}{*}{\!\!\!WMMSE\!\!\!} & \multicolumn{12}{c|}{DMPNN ($p_{\text{train}}=0.8$)}                                                                                                                                                                                                                                                                                                                       \\ \cline{4-15}
                     &                      &                        & \multicolumn{4}{c||}{$p_{\text{test}}=0.5$}                                                                                                   & \multicolumn{4}{c||}{$p_{\text{test}}=0.7$}                                                                          & \multicolumn{4}{c|}{Complete graph}                                                                          \\ \cline{4-15}
                     &                      &                        & $\!\!M=3\!\!$                           & $\!\!M=5\!\!$                           & $\!\!M=10\!\!$                          & $\!\!M=15\!\!$                          & $\!\!M=3\!\!$                           & $\!\!M=5\!\!$                           & $\!\!M=10\!\!$                          & $\!\!M=15\!\!$ & $\!\!M=3\!\!$                           & $\!\!M=5\!\!$                           & $\!\!M=10\!\!$                          & $\!\!M=15\!\!$ \\ \hline
3                    & 2.909                & 2.760                  & \textbf{2.772} & 2.767                           & 2.734                           & 2.730                           & \textbf{2.839} & 2.832                           & 2.823                           & 2.824  & \textbf{2.903} & 2.901                           & 2.899                           & 2.901  \\ \hline
5                    & 3.338                & 3.190                  & 3.207                           & \textbf{3.220} & \textbf{3.220} & 3.198                           & \textbf{3.286} & 3.284                           & 3.291                           & 3.286  & 3.345                           & \textbf{3.350} & 3.344                           & 3.346  \\ \hline
7                    & 3.654                & 3.518                  & 3.491                           & 3.505                           & \textbf{3.507} & 3.499                           & 3.580                           & \textbf{3.591} & \textbf{3.591} & 3.586  & 3.654                           & 3.665                           & \textbf{3.657} & 3.656  \\ \hline
9                    & 3.877                & 3.758                  & 3.711                           & 3.733                           & 3.727                           & \textbf{3.734} & 3.814                           & \textbf{3.831} & 3.826                           & 3.828  & 3.913                           & \textbf{3.936} & 3.922                           & 3.917  \\ \hline
\end{tabular}\label{table:table2b}
}
\label{table:table2}
\end{table*}
\endgroup

The impact of the hidden layer dimension is investigated in Table \ref{table:table1} for $p_{\text{train}}=0.6$. 
An ordered pair $(U,S)$ stands for $U$-neuron hidden layers used for FNNs in \eqref{eq:fnnm}-\eqref{eq:fnnd} and $S$-neuron hidden layers used for the GRU in \eqref{eq:grus}.
The best performance is marked in boldface for each $p_{\text{test}}$. The average sum rate performance is enhanced by the increase of $U$ and $S$, while the computational complexity of the real-time inference grows. As a consequence, $(100,50)$ is chosen for The DMPNN since the performance improvement becomes clipped at this combination.

The message dimension affects the signaling overhead significantly. Although the choice of a small value for $M$ simplifies the system design, it generally degrades the learning performance. Table \ref{table:table2} shows the average sum rate of the DMPNN with various values of $M$. For all simulated configurations of $M$ and $p_{\text{train}}$, the DMPNN exhibits almost identical performance to the WMMSE algorithm for  $p_{\text{test}}=0.5$.
Large message dimensions, such as $M=10$ and $15$, are necessary for training the DMPNN at $p_{\text{train}}=0.6$ (Table \ref{table:table2a}), whereas small values of $M$ suffice for the case of $p_{\text{train}}=0.8$ (Table \ref{table:table2b}). The distributed optimization over less-connected backhaul configurations is generally more difficult. Thus, nodes exchange high-dimensional messages for an improved performance with limited coordination. In the case of $p_{\text{train}}=0.8$, the test performance decreases for $M \geq 5$. This indicates that the DMPNN with $M=15$ has an excessively large number of parameters and is likely to overfit simple configurations with high $p_{\text{train}}$. From these results, the message dimension is set to $M=10$ afterwards, since this shows consistently good performance over various values of $p_{\text{test}}$.

\begin{figure}
\centering
    \includegraphics[width=.5\linewidth]{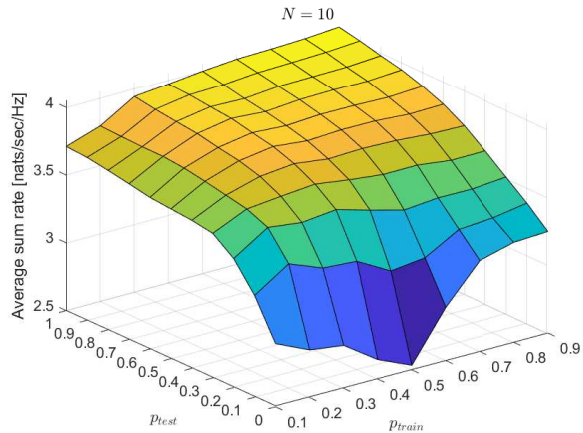}
    \caption{Average sum rate for $N=10$ with various combinations of $p_{\text{train}}$ and $p_{\text{test}}$.}
    \label{simul:fig3}
\end{figure}

\begin{figure}
\centering
    \subfigure[Examples of social graphs with $N=5$.]{
        \includegraphics[width=.32\linewidth]{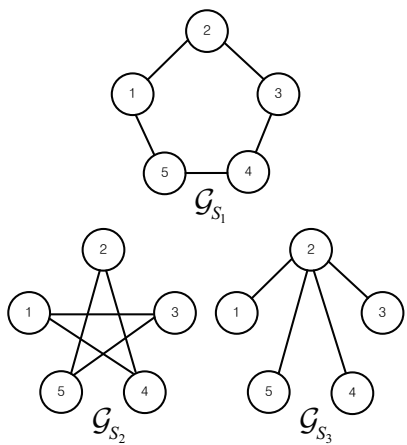}\label{simul:fig2a}
    }
    \subfigure[Average sum rate as a function of $t$.]{
        \includegraphics[width=.5\linewidth]{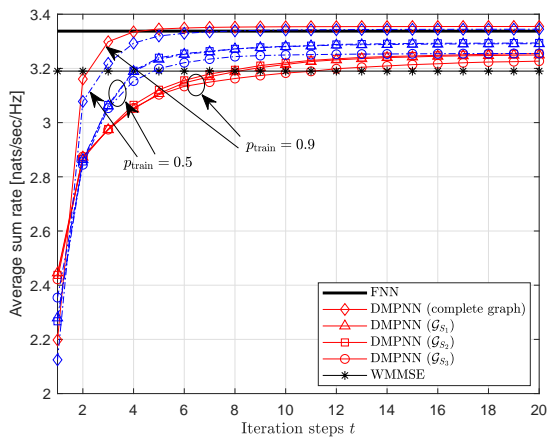}\label{simul:fig2b}
    }
    \caption{Average sum rate with fixed social graphs.}
    \label{simul:fig2}
\end{figure}

\begin{figure}
\centering
\includegraphics[width=.5\linewidth]{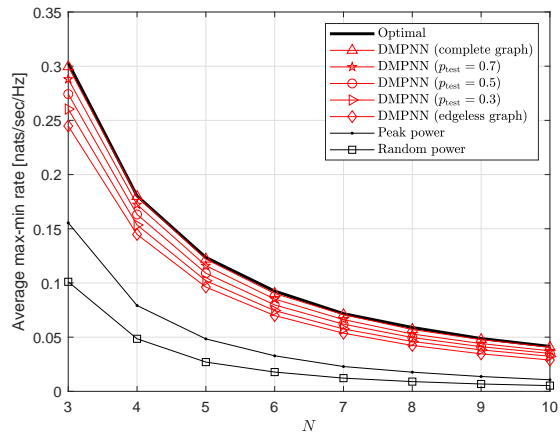}
\caption{Average max-min rate as a function of $N$ with $p_{\text{train}}=0.7$.}
\label{simul:fig5}
\end{figure}

Fig. \ref{simul:fig3} shows how well the DMPNN performs in the range of  $p_{\text{test}}\in\{0,0.1,\cdots,0.9,1\}$ for $N=8$ when trained at $p_{\text{train}}\in\{0.1,0.2,\cdots,0.9\}$. One can see that the choice of $p_{\text{train}}$ equal to $p_{\text{test}}$, i.e., $p_{\text{train}}=p_{\text{test}}$, is not necessarily the most efficient for a certain range of $p_{\text{test}}$. In particular, the DMPNNs trained at $p_{\text{train}}=0.7$, $0.8$, and $0.9$ exhibit better performance than trained at other probability values. This can be justified as follows: While trained at small $p_{\text{train}}$, the DMPNN does not observe sufficient samples of messages $\mathbf{m}_{ij}^{[t]}$ and their associated aggregates $\mathbf{c}_{ji}^{[t]}$ since the average neighborhood population is small in Erdos-Renyi graphs with small $p_{\text{train}}$. Thus, the DMPNN may not learn exact computation rules of message generation and combination in such graph configurations. By contrast, the DMPNN is able to learn an efficient information sharing strategy in high $p_{\text{train}}$ regimes, thereby leading to the performance improvement over a wide range of $p_{\text{test}}$.

Fig. \ref{simul:fig2a} enumerates several $5$-node backhaul configurations where the performance of the DMPNN trained over Erdos-Renyi graphs is examined. Note that $\mathcal{G}_{S_{2}}$ is a permutation of $\mathcal{G}_{S_{1}}$ with $\pi(1)=1$, $\pi(2)=3$, $\pi(3)=5$, $\pi(4)=2$, and $\pi(5)=4$. It is thus expected that the DMPNN behaves similarly under two graphs. Besides, $\mathcal{G}_{S_{3}}$ corresponds to the case where node 2 acts as a central coordinator for the distributed optimization. Fig. \ref{simul:fig2b} presents the convergence behaviors for the first 20 iterations with various values of $p_{\text{train}}$. The convergence of the sum rate objective improves as $t$ grows, implying that the weighted sum utility objective $\mathcal{F}(\mathbf{\Theta})$ is properly designed for the guarantee of the convergence. Upon the convergence of messages, the DMPNN outperforms the WMMSE algorithm consistently over all tested graph configurations. Fig. \ref{simul:fig2a} shows the improvement of the convergence rate with the complete graph over non-complete graph cases. As expected, two permutations $\mathcal{G}_{S_{1}}$ and $\mathcal{G}_{S_{2}}$ exhibit an identical convergence behavior, verifying the permutation equivariance of the DMP inference in Proposition \ref{prop:prop01}. The performance tested over $\mathcal{G}_{S_{3}}$ is degraded as compared to the cases with $\mathcal{G}_{S_{1}}$ and $\mathcal{G}_{S_{2}}$ since a node in $\mathcal{G}_{S_{3}}$ has, on average,  less number of neighbors than those in $\mathcal{G}_{S_{1}}$ and $\mathcal{G}_{S_{2}}$. The design of training techniques for the information sharing and the distributed optimization over $\mathcal{G}_{S_{1}}$ becomes intricate. Furthermore, the convergence of the DMPNN is affected by the probability $p_{\text{train}}$. To be precise, the DMPNN trained at $p_{\text{train}}=0.9$ performs well for the complete social graph, whereas non-complete counterparts in Fig. \ref{simul:fig2a} show good performance with $p_{\text{train}}=0.5$.


\subsection{Minimum Rate Maximization}

\begin{figure}
\centering
    \includegraphics[width=.5\linewidth]{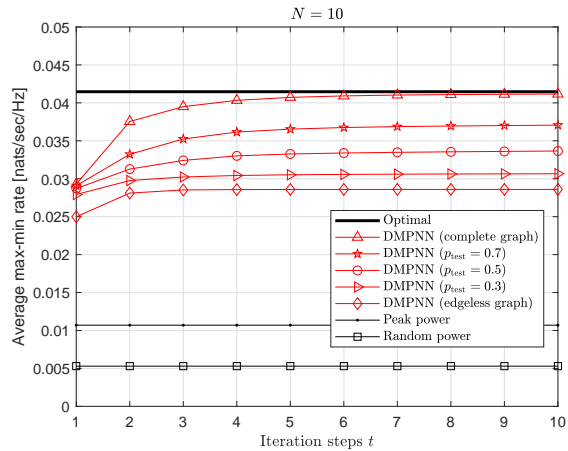}
    \caption{Average max-min rate as a function of $t$ with $N=10$ and $p_{\text{train}}=0.9$.}
    \label{simul:fig6}
\end{figure}

The maximin formulation in (P2) is considered with a deep-layered DMPNN architecture where the FNNs in \eqref{eq:fnnm}-\eqref{eq:fnnd} are constructed with four layers and $150$ hidden-layer dimension.\footnote{The SGD update rules are replaced by the subgradient method \cite{Boyd:04} at nondifferentiable points where gradients are unavailable.} The maximum number of the iterations is set to $T=10$. Fig.~\ref{simul:fig5} depicts the average maximin objective value of the DMPNN trained over Erdos-Renyi graphs with $p_{\text{train}}=0.7$ for various values of $N$. A globally optimal algorithm \cite{DCai:12}, which operates over the complete social graph for an iterative computation of the power control solution, is compared as a benchmark. The DMPNN tested over the complete graph shows nearly the optimal performance over all simulated node populations. As seen from the max-sum problem in (P1), the performance of the DMPNN remains competitive even in case of missing backhaul cooperation, i.e., over the edgeless graph.

\begin{figure}
\centering
    \includegraphics[width=.5\linewidth]{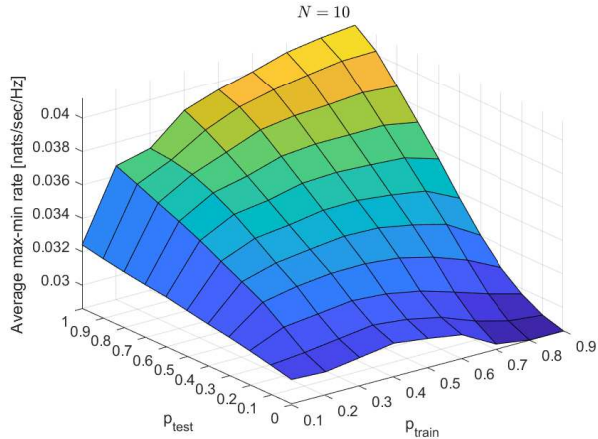}\label{simul:fig7b}
    \caption{Average max-min rate for $N=10$ with various combinations of $p_{\text{train}}$ and $p_{\text{test}}$.}
    \label{simul:fig7}
\end{figure}

\begin{figure}
\centering
    \subfigure[Max-sum problem.]{
        \includegraphics[width=.45\linewidth]{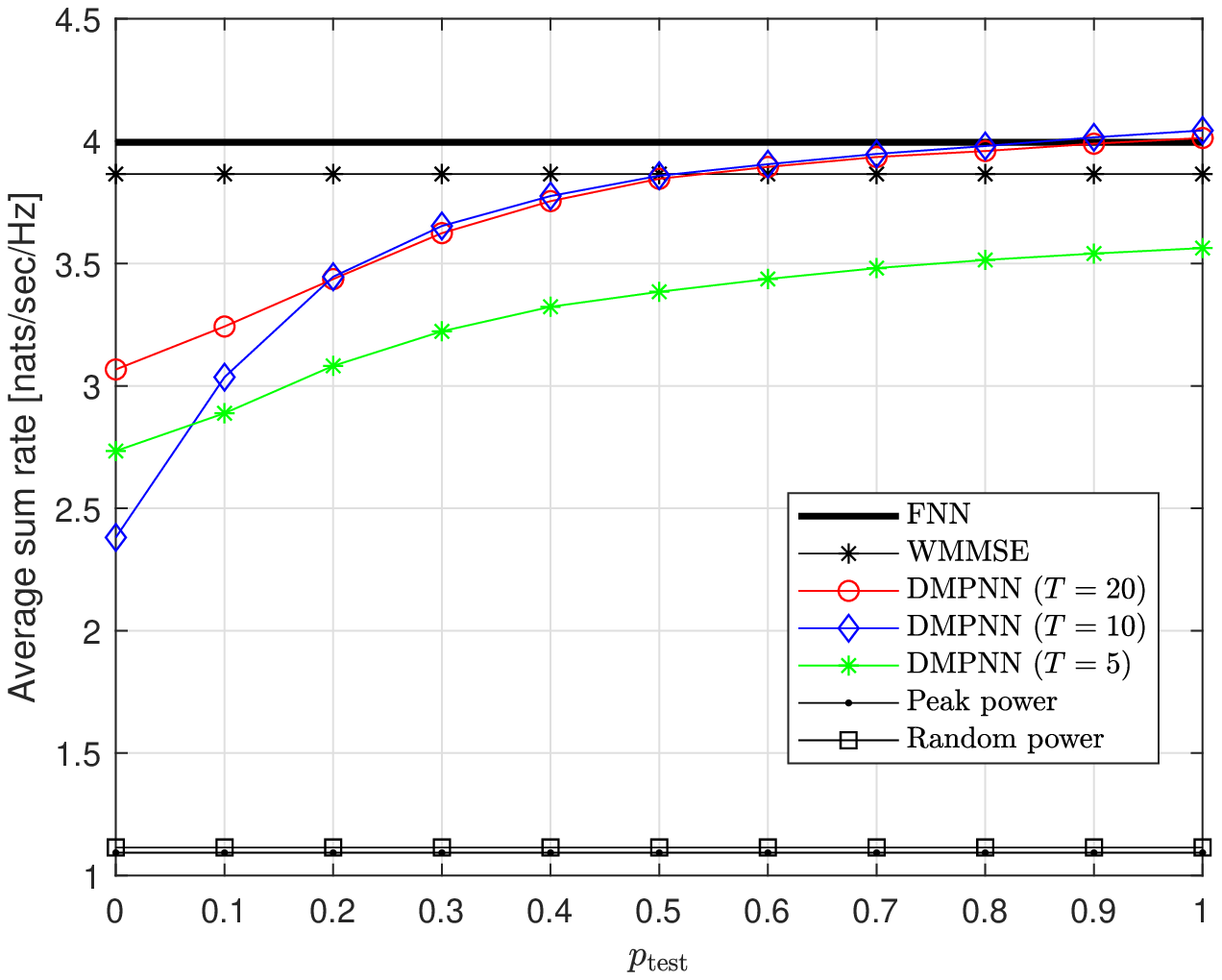}\label{fig:figR2C4a}
    }
    \subfigure[Max-min problem.]{
        \includegraphics[width=.45\linewidth]{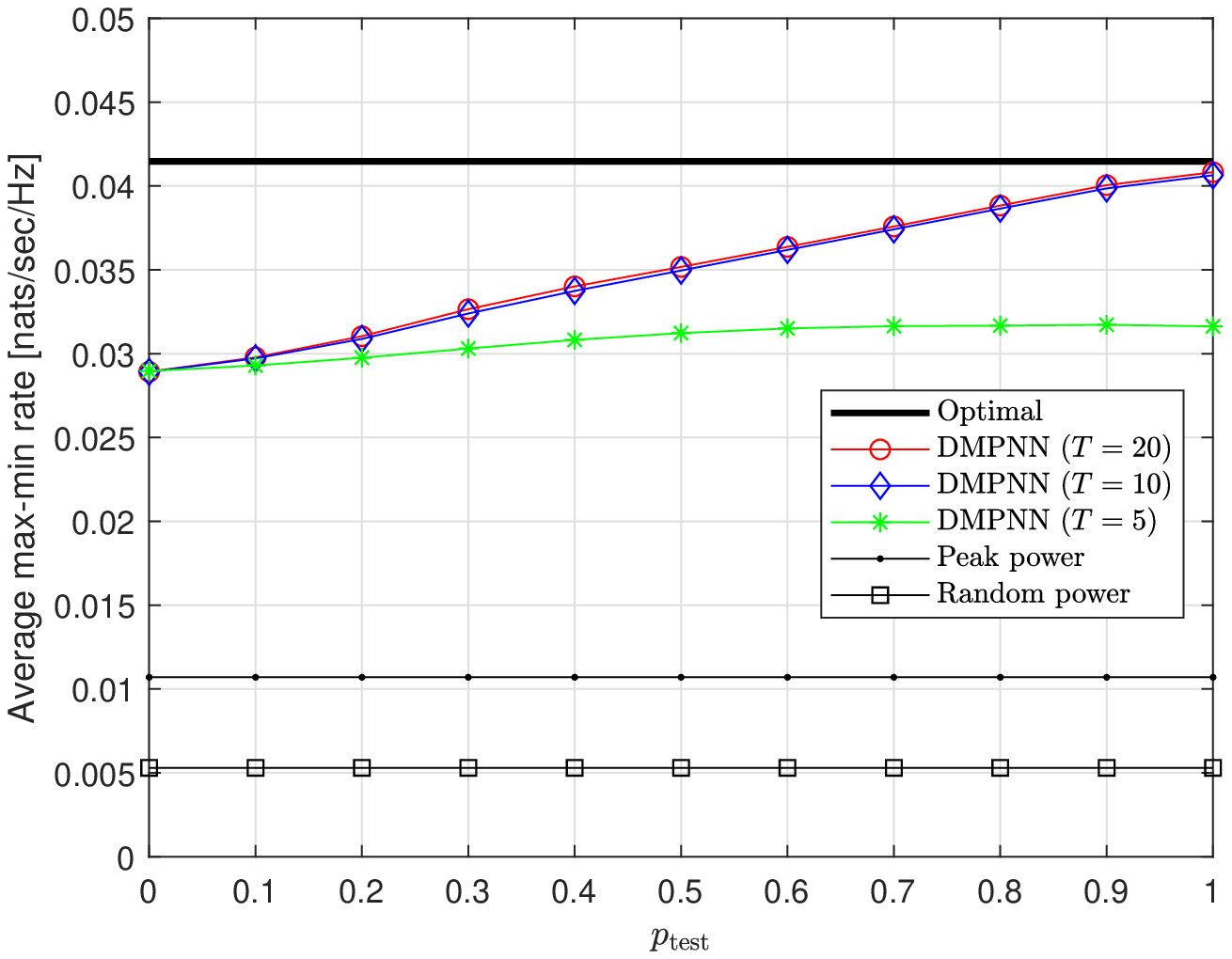}\label{fig:figR2C4b}
    }
    \caption{Average sum rate and max-min rate performance as a function of $p_{\text{test}}$ with $p_{\text{train}}=0.7$.}
    \label{fig:figR2C4}
\end{figure}

Fig. \ref{simul:fig6} illustrates the average maximin rate with $p_{\text{train}}=0.9$ for $N=5$ and $10$. Regardless of the node population and the test graph setup, the forwardpass evaluation converges within 10 iterations. Fig. \ref{simul:fig7} presents the maximin performance for various combinations of $p_{\text{train}}$ and $p_{\text{test}}$, showing similar results with the max-sum case in Fig. \ref{simul:fig3}. Therefore, $p_{\text{train}}=0.8$ and $0.9$ are suggested for the performance improvement over random social interactions.
Finally, Fig. \ref{fig:figR2C4} shows the impact of $T$ by demonstrating the performance of max-sum problem (P1) (Fig. \ref{fig:figR2C4a}) and max-min problem (P2) (Fig. \ref{fig:figR2C4b}) with respect to $p_{\text{test}}$. A large value of $T$ deepens the DMPNN architecture that includes additional powerful DNNs and increases training computations. The average sum rate enhances with $T=20$, in particular, at low $p_{\text{test}}$ regimes, whereas the minimum rate reveals only a slight improvement after $T=10$. Thus, the choice of $T=10$ is computation-efficient for the minimum rate maximization task (P2).

\section{Conclusions}\label{sec:sec7}
This work studies a novel DL framework that addresses universal distributed optimization over random networks with random configurations of node population and backhaul connection topology. This universal formulation has not been approached very well with conventional optimization and DL techniques for their typical rigid computation structures. To handle this difficulty, a DMP inference algorithm that handles arbitrary networking configurations is proposed first. Subsequently, a model-driven DMPNN framework is developed to learn the DMP inference. Numerical results verify that the proposed framework efficiently can address distributed max-sum and max-min power control tasks over varying network topology in the IFC. For a future work, the identification of its decentralized learning technique is worthwhile to pursue.

\begin{appendices}
\section{Proof of Proposition \ref{prop:prop0}}\label{app:appAA}
It is remarked that the information made available at each node $i$ is the concatenation of local observation $\mathbf{a}_{ii}$ and external data set $\{\mathbf{a}_{ji}:j\in\mathcal{N}_{P}(i)\}$ sensed passively from the physical neighborhood. Node $i$ is informed of the collection of computation results  $\{\mathbf{x}_{j}:j\in\mathcal{N}_{S}(i)\}$ via backhaul cooperations with the social neighborhood. Let $\mathcal{D}_{i}(\cdot)$ denote the optimal distributed decision operator of node $i$ with objective function $f(\cdot)$ and all local information. 
By definition, the solution $\mathbf{x}_{i}$ associated with node $i$ is expressed as
\begin{align}
\mathbf{x}_{i}=\mathcal{D}_{i}(\{\mathbf{x}_{j}:j\in\mathcal{N}_{S}(i)\},\{\mathbf{a}_{ji}:j\in\mathcal{N}_{P}(i)\},\mathbf{a}_{ii}).\label{eq:di}
\end{align}
For simplicity, the permutated index of node $i$ is designated as $k\triangleq\pi(i)$. The solution $\mathbf{x}_{k}^{\pi}$ associated with node $k\in\mathcal{V}^{\pi}$ is expressed as
\begin{align}
\mathbf{x}_{k}^{\pi}&=\mathcal{D}_{k}
(\{\mathbf{x}_{j}^{\pi}:j\in\mathcal{N}_{S}^{\pi}(k)\},\{\mathbf{a}_{jk}^{\pi}:j\in\mathcal{N}_{P}^{\pi}(k)\},\mathbf{a}^{\pi}_{kk})\nonumber\\
    &=\mathcal{D}_{k}(\{\mathbf{x}_{j}:j\in\mathcal{N}_{S}(i)\},\{\mathbf{a}_{ji}:j\in\mathcal{N}_{P}(i)\},\mathbf{a}_{ii}),\label{eq:dpi}
\end{align}
where $\mathcal{N}_{S}^{\pi}(i)\triangleq\{j:(i,j)\in\mathcal{E}_{S}^{\pi}\}$ and $\mathcal{N}_{P}^{\pi}(i)\triangleq\{j:(i,j)\in\mathcal{E}_{P}^{\pi}\}$ are the neighborhoods of node $i$ in the permuted graphs $\mathcal{G}_{S}^{\pi}$ and $\mathcal{G}_{P}^{\pi}$, respectively, and \eqref{eq:dpi} results from the fact that $\mathcal{N}_{S}^{\pi}(k)=\mathcal{N}_{S}(\pi^{-1}(k))=\mathcal{N}_{S}(i)$, $\mathcal{N}_{P}^{\pi}(k)=\mathcal{N}_{P}(\pi^{-1}(k))=\mathcal{N}_{P}(i)$, and $\mathbf{a}_{jk}^{\pi}=\mathbf{a}_{\pi^{-1}(j)\pi^{-1}(k)}=\mathbf{a}_{\pi^{-1}(j)i}$.

Note that $\mathcal{D}_{i}(\cdot)$ in \eqref{eq:di} and $\mathcal{D}_{k}(\cdot)=\mathcal{D}_{\pi(i)}(\cdot)$ in \eqref{eq:dpi} have the same arrangements of the input. Thus, it suffices to show the existence of the common optimal rule such that $\mathcal{D}_{i}(\cdot)=\mathcal{D}_{\pi(i)}(\cdot)$. It is easily seen since $\mathcal{D}_{i}(\cdot)$ and $\mathcal{D}_{\pi(i)}(\cdot)$ evaluate $f(\mathbf{a},\mathbf{x})$ and $f(\mathbf{a}^{\pi},\mathbf{x}^{\pi})$, respectively, satisfying
\begin{align}
f(\mathbf{a}^{\pi},\mathbf{x}^{\pi}) =&f(\{\mathbf{a}_{i}^{\pi}:i\in\mathcal{V}^{\pi}\},\{\mathbf{x}_{i}^{\pi}:i\in\mathcal{V}^{\pi}\})\nonumber\\
=&f(\{\mathbf{a}^{\pi}_{\pi(i)}:i\in\mathcal{V}\},\{\mathbf{x}_{\pi(i)}^{\pi}:i\in\mathcal{V}\})=f(\mathbf{a},\mathbf{x}).
\end{align}
The resulting outputs for $\mathcal{D}_{i}(\cdot)$ and $\mathcal{D}_{\pi(i)}(\cdot)$ are identical, i.e., $\mathbf{x}_{i}=\mathbf{x}_{\pi(i)}^{\pi}$ for all $i\in\mathcal{V}$. Hence, there exists the optimal distributed solution satisfying \eqref{eq:prop0} for the permutations $\mathcal{G}_{S}^{\pi}$ and $\mathcal{G}_{P}^{\pi}$. \qqed

\section{Derivation of DMP Operations}\label{app:appB}
We begin with the optimal distributed solution computation policy in \eqref{eq:di}. According to Proposition \ref{prop:prop0}, subscript $i$ is removed from the decision operator $\mathcal{D}_{i}(\cdot)$ for the permutation invariance property. It thus follows that \eqref{eq:di} is rewritten as
\begin{align}
\mathbf{x}_{i}=\mathcal{D}(\{\mathbf{x}_{j}:j\in\mathcal{N}_{S}(i)\},\{\mathbf{a}_{ji}:j\in\mathcal{N}_{P}(i)\},\mathbf{a}_{ii}).\label{eq:xi}
\end{align}
A distributed solution of each node $i$ is normally achieved by iterative computations with adjacent nodes' local decisions $\{\mathbf{x}_{j}:j\in\mathcal{N}_{S}(i)\}$. An iterative form of the local decision in \eqref{eq:xi} is reexpressed at the $t$-th iteration as
\begin{align}
\mathbf{x}_{i}^{[t]}&=\mathcal{D}(\mathbf{x}_{i}^{[t-1]},\{\mathbf{x}_{j}^{[t-1]}:j\in\mathcal{N}_{S}(i)\},\{\mathbf{a}_{ji}:j\in\mathcal{N}_{P}(i)\},\mathbf{a}_{ii})\nonumber\\
                    &=\mathcal{D}(\mathbf{x}_{i}^{[t-1]},\{\mathbf{c}_{ji}^{[t]}:j\in\mathcal{N}(i)\},\mathbf{a}_{ii}), \label{eq:xit}
\end{align}
where $\mathcal{N}(i)\triangleq\mathcal{N}_{S}(i)\cup\mathcal{N}_{P}(i)$ and $\mathbf{c}_{ji}^{[t]}$ is the collection of the information available at node $i$ originating from neighboring node $j$ given as
\begin{align}
\mathbf{c}_{ji}^{[t]}\label{eq:cjit}
=\begin{cases}
    (\mathbf{x}_{j}^{[t-1]},\mathbf{a}_{ji}) &\text{if}~j\in\mathcal{N}_{S}(i)\cap\mathcal{N}_{P}(i),\\
    \mathbf{x}_{j}^{[t-1]} &\text{if}~j\in\mathcal{N}_{S}(i)\cap\mathcal{N}_{P}^c(i),\\
    \mathbf{a}_{ji} &\text{if}~j\in\mathcal{N}_{S}^c(i)\cap\mathcal{N}_{P}(i),
\end{cases}
\end{align}
since the determination of $\mathbf{x}_{i}^{[t]}$ can benefit from the knowledge of $\mathbf{x}_{i}^{[t-1]}$ which is obviously available at node $i$.

Furthermore, 3-tuple inputs of \eqref{eq:xit} depend only on the information available at node $i$. This indicates that such a triplet can be defined in a state $\mathbf{s}_{i}^{[t]}$ as
\begin{align}
\mathbf{s}_{i}^{[t]}=(\mathbf{x}_{i}^{[t-1]},\{\mathbf{c}_{ji}^{[t]}:j\in\mathcal{N}(i)\},\mathbf{a}_{ii}).\label{eq:sit1}
\end{align}
Thus, the decision operation in \eqref{eq:xit} can be reexpressed in a single-input function as
\begin{align}
\mathbf{x}_{i}^{[t]}=\mathcal{D}(\mathbf{x}_{i}^{[t-1]},\{\mathbf{c}_{ji}^{[t]}:j\in\mathcal{N}(i)\},\mathbf{a}_{ii})=\mathcal{D}(s_{i}^{[t]}).\label{eq:xit2}
\end{align}
By plugging \eqref{eq:xit2} into \eqref{eq:sit1}, the state in \eqref{eq:sit1} can also be expressed in a recursive form as
\begin{align}
\mathbf{s}_{i}^{[t]}=(\mathcal{D}(\mathbf{s}_{i}^{[t-1]}),\{\mathbf{c}_{ji}^{[t]}:j\in\mathcal{N}(i)\},\mathbf{a}_{ii}).\label{eq:sit2}
\end{align}
Note that this provides a state update rule represented in state update operation $\mathcal{S}(\cdot)$ as
\begin{align}
\mathbf{s}_{i}^{[t]}&=\mathcal{S}(\mathbf{s}_{i}^{[t-1]},\{\mathbf{c}_{ji}^{[t]}:j\in\mathcal{N}(i)\},\mathbf{a}_{ii}).\label{eq:sit3}
\end{align}

Meanwhile, local decision $\mathbf{x}_{j}^{[t-1]}$ in \eqref{eq:cjit} is unavailable at node $i$ since it is not explicitly transferred out of node $j$. Therefore, the message $\mathbf{m}_{ij}^{[t]}\in\mathbb{R}^{M}$ transmitted from node $i$ to node $j$ is required to convey sufficient information required for the optimal distributed decision. The message $\mathbf{m}_{ij}^{[t]}$ encodes $\mathbf{s}_{i}^{[t-1]}$ since $\mathbf{x}_{j}^{[t-1]}$ depends only on $\mathbf{s}_{i}^{[t-1]}$. To generate $\mathbf{m}_{ij}^{[t]}$ dedicated to node $j$, local information $\mathbf{a}_{ji}$ can be used as additional input to complement $\mathbf{s}_{i}^{[t-1]}$. Thus, using the message generation operator $\mathcal{M}(\cdot)$, $\mathbf{m}_{ij}^{[t]}$ can be written as $\mathbf{m}_{ij}^{[t]}=\mathcal{M}(\mathbf{s}_{i}^{[t-1]},\tilde{\mathbf{a}}_{ji})$
where $\tilde{\mathbf{a}}_{ji}$ in \eqref{eq:aji_tilde} is defined depending on the availability of the connection between node $i$ and node~$j$.

Noted that, based on the message formulation \eqref{eq:cjit} and \eqref{eq:sit1}, $\mathbf{s}_{i}^{[t-1]}$ has $\mathbf{c}_{ji}^{[t]}$ as an input and, in turn, requires the knowledge about $\mathbf{x}_{j}^{[t-1]}$, which is available from the dedicated message $\mathbf{m}_{ji}^{[t]}$. Thus, with the combination operator $\mathcal{C}(\cdot)$, $\mathbf{c}_{ji}^{[t]}$ in \eqref{eq:cjit} is recast into
\begin{align}
\mathbf{c}_{ji}^{[t]}\label{eq:cjit2}
=\begin{cases}
    \mathcal{C}(\mathbf{m}_{ji}^{[t]},\mathbf{a}_{ji}) &\text{if}~j\in\mathcal{N}_{S}(i)\cap\mathcal{N}_{P}(i)\\
    \mathcal{C}(\mathbf{m}_{ji}^{[t]},\mathbf{0}_{K_{2}}) &\text{if}~j\in\mathcal{N}_{S}(i)\cap\mathcal{N}_{P}^c(i)\\
    \mathcal{C}(\mathbf{0}_{M},\mathbf{a}_{ji}) &\text{if}~j\in\mathcal{N}_{S}^c(i)\cap\mathcal{N}_{P}(i)
\end{cases}
=\mathcal{C}(\tilde{\mathbf{m}}_{ji}^{[t]},\tilde{\mathbf{a}}_{ji}),
\end{align}
where $\tilde{\mathbf{m}}_{ji}^{[t]}$ and $\tilde{\mathbf{a}}_{ji}$ are defined in \eqref{eq:mjit_tilde} and \eqref{eq:aji_tilde}, respectively.
Furthermore, the collection of $\{\mathbf{c}_{ji}^{[t]}\}$ is represented with the aggregation operator $\mathcal{A}(\cdot)$, and the corresponding output is denoted by $\mathbf{c}_{i}^{[t]}$ as  \eqref{eq:cita}. The new definition is used to simplify the state update operator \eqref{eq:sit3} into \eqref{eq:sita} as $\mathbf{s}_{i}^{[t]}=\mathcal{S}(\mathbf{s}_{i}^{[t-1]},\mathbf{c}_{i}^{[t]},\mathbf{a}_{ii})$.
Therefore, all DMP operators in \eqref{eq:mijta}-\eqref{eq:xita} are obtained. \qqed

\end{appendices}
\begingroup
\renewcommand{\baselinestretch}{1.4}
\bibliographystyle{IEEEtran}
\bibliography{arXiv}
\endgroup

\end{document}